\documentclass{article}

\usepackage[utf8]{inputenc}
\usepackage[T1]{fontenc}
\usepackage[english]{babel}
\usepackage{amsmath,amssymb,amsthm,mathtools}
\usepackage{tikz-cd}
\usepackage{graphicx} 
\usepackage[margin=1.in]{geometry}
\usepackage[shortlabels]{enumitem}
\usepackage{todonotes}
\usepackage[bibstyle=numeric, backend=biber, sorting=none, citestyle=numeric-comp, giveninits,url=false,maxbibnames=5]{biblatex} 
\usepackage{csquotes}\MakeOuterQuote"
\usepackage{soul}\setstcolor{red}
\usepackage[colorlinks,citecolor=blue,linkcolor=blue]{hyperref}
\usepackage[capitalize]{cleveref}
\usepackage{comment}
\usepackage{booktabs}

\DeclareUnicodeCharacter{0141}{\Lbar{}}

\newtheorem{theorem}{Theorem}
\newtheorem{lemma}[theorem]{Lemma}
\newtheorem{corollary}[theorem]{Corollary}
\newtheorem{proposition}[theorem]{Proposition}

\newtheorem{definition}[theorem]{Definition}

\newtheorem{introtheorem}{Theorem} 

\newtheorem{introproposition}[introtheorem]{Proposition} 

\newtheorem{introlemma}[introtheorem]{Lemma}

\theoremstyle{definition}

\newtheorem*{problem*}{Problem}
\newtheorem*{assumption*}{Assumption}

\newtheorem{remark}[theorem]{Remark}
\newtheorem*{warning*}{Warning}

\DeclareMathOperator{\mult}{MD}

\newcommand{\ip}[2]{\langle #1,#2\rangle}

\newcommand{\ketbra}[2]{|#1\rangle\langle#2|}

\newcommand{\kettbra}[1]{\ketbra{#1}{#1}}

\DeclareMathOperator{\supp}{supp}
\newcommand{\norm}[1]{\lVert #1\rVert}
\newcommand{\oo}{\infty}
\newcommand{\ox}{\otimes}
\newcommand{\mc}{\mathcal}
\newcommand{\eps}{\varepsilon}

\newcommand{\II}{{\mathrm{II}}}
\newcommand{\I}{{\mathrm{I}}}

\DeclareMathOperator{\tr}{tr}

\newcommand{\hide}[1]{}

\def\B{{\mc B}}
\def\CC{{\mathbb C}}

\def\H{{\mc H}}

\def\K{{\mathcal K}}

\def\RR{{\mathbb R}}

\def\NN{{\mathbb N}}

\newcommand{\proj}{\mathrm{Proj}}

\DeclareMathOperator{\lin}{span}
\DeclareMathOperator{\id}{id}

\newcommand*{\1}{\text{\usefont{U}{bbold}{m}{n}1}}
\newcommand{\placeholder}[0]{{\,\cdot\,}}

\newcommand{\Lbar}{\L{}}

\usetikzlibrary{patterns,positioning,arrows.meta,decorations.pathreplacing,
                decorations.pathmorphing,fit,calc}
\definecolor{Acol}{RGB}{190,206,190}
\definecolor{Adark}{RGB}{62,102,76}
\definecolor{Bcol}{RGB}{255,229,201}
\definecolor{Bdark}{RGB}{198,116,54}
\definecolor{Gapcol}{RGB}{228,228,233}
\definecolor{Gapdark}{RGB}{112,112,124}

\title{
Quantum steering is equivalent to state-preserving conditional expectations
}
\author{
Lauritz van Luijk$^{1,2}$, Amine Marrakchi$^3$, Tobias Osborne$^4$, \\ Alexander Stottmeister$^4$, and Henrik Wilming$^4$
}

\date{
{\small
$^1$Perimeter Institute for Theoretical Physics, Waterloo, Ontario, Canada\\[2pt]
$^2$Institute for Quantum Computing, Waterloo, Ontario, Canada\\[2pt]
$^3${Unité de Mathématiques Pures et Appliquées, CNRS - ENS Lyon, Lyon, France} \\[2pt]
$^4$Leibniz Universität Hannover, Institut für Theoretische Physik, Appelstraße 2, 30167 Hannover, Germany}\\[11pt]
\today}

\begin{document}

\maketitle

\begin{abstract}
    In systems with infinitely many degrees of freedom, fundamental results from quantum information theory can fail.
    An important example is the uniqueness of purifications:
    Even when two subsystems, described by commuting von Neumann algebras $A$ and $B$, are tomographically complete, purifications of a state on $A$ need not be related by unitaries in $B$.
    It was recently shown that this occurs precisely when Haag duality fails, i.e., when the commutant $B'$ is strictly larger than $A$.
    This raises the question of which fundamental entanglement properties survive in such a setting. 
    We show that, for a pure global state, the ability to \emph{steer} any ensemble decomposition of the marginal state on $A$ by measurements on $B$ is equivalent to the existence of a state-preserving conditional expectation from $B'$ onto $A$.
    This establishes a direct connection between quantum steering and subfactor theory.
    The key observation is that steering is equivalent to the existence of extensions of ensemble decompositions from $A$ to $B'$.
    Working with general Jordan algebras, we prove that unital positive maps have state-preserving left inverses if and only if ensemble decompositions can be lifted.
    For the inclusion $A\hookrightarrow B'$, a left inverse is precisely a conditional expectation, yielding the characterization above.
\end{abstract}

\tableofcontents

\clearpage

\section{Introduction and Overview}\label{sec:intro}

Entanglement in quantum systems with infinitely many degrees of freedom is receiving increased attention in recent time.
To obtain mathematically precise results, the subsystems in such systems are typically modeled by commuting von Neumann algebras $A,B$ on a joint Hilbert space $\H$.
Recent results established clear operational interpretations of basic operator algebraic results in terms of entanglement theory \cite{summersMaximalViolationBells1987,verch_distillability_2005,keylEntanglementHaagdualityType2006,crannStateConvertibilityNeumann2020,van_luijk_embezzlement_2024,van_luijk_relativistic_2024,vanluijkPureStateEntanglement2025,vanluijkUniquenessPurificationsEquivalent2026}, such as the type classification of factors or the notion of Haag duality ($A = B'$, $B'$ denoting the commutant of $B$), which was first introduced in algebraic quantum field theory \cite{haag_local_1996}, but recently also plays an important role in the rigorous understanding of topological quantum order \cite{naaijkensAnyonsInfiniteQuantum2012,naaijkensHaagDualityDistal2012,fiedlerHaagDualityKitaevs2015,ogataDerivationBraidedCtensor2022,jonesLocalTopologicalOrder2025a,ogataHaagDuality2D2025,naaijkensLocalTopologicalOrder2026} and in the study of non-invertible symmetries \cite{shaoAdditivityHaagDuality2025,harlowDisjointAdditivityLocal2025a}.

In fact, Haag duality is equivalent to a central property in quantum information theory, namely the uniqueness of purifications \cite{vanluijkUniquenessPurificationsEquivalent2026}: If a state vector (i.e., unit vector) $\Omega\in \H$ induces a state $\omega$ on $A$, then every other purification $\Psi \in \H$ of $\omega$ can be approximated by acting with a unitary $u\in B$ on $\Omega$ if and only if $A=B'$.  
Uniqueness of purifications (Haag duality) is central to pure state entanglement theory. 
It is hence interesting to ask which operational properties of pure state entanglement remain without it.

\begin{figure}
	\center
	\begin{tikzpicture}
	  \node[anchor=south west,inner sep=0] (f1img) at (0,0)
	        {\includegraphics[width=4cm]{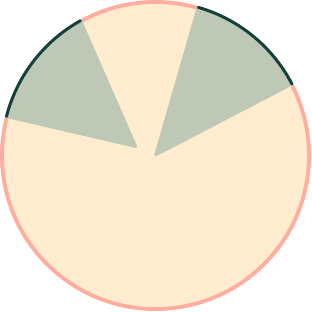}};
	  \begin{scope}[x={(f1img.south east)},y={(f1img.north west)}]
	    \node at (0.229,0.704) {$A$};
	    \node at (0.697,0.746) {$A$};
	    \node at (0.498,0.250) {$B$};
	  \end{scope}
	\end{tikzpicture}
   { {}\\ \vspace{0.5cm} }
     \begin{tabular}{@{}llccl@{}}
    & region $A$ & Haag duality & preserves $\omega$? & obstruction \\
    \midrule
    surface code (bulk)
      & two disjoint cones     & fails & \checkmark\ ground state & ---\\
    rational CFT (boundary)
      & two disjoint intervals & fails & $\times$\ vacuum
      & $\Omega$ cyclic for $A$\\
    \bottomrule
  \end{tabular}
    \label{fig:disjoint-cones}
	\caption{Examples; Bulk: A two-dimensional topologically ordered system is divided into a region $A$ consisting of two disjoint infinite cones and their complement $B$. Boundary: A completely rational 1+1-dimensional conformal field theory is divided into a region $A$ consisting of two disjoint intervals and their complement $B$.
    Haag duality fails in both settings and a normal conditional expectation $B'\to A$ exists, so by \cref{introthm:main} the relevant question is which state is preserved. For the surface code it is the ground
    state; for a completely rational chiral conformal net it cannot be the vacuum, because of the Reeh--Schlieder property.}
   
\end{figure}

Interestingly, Haag duality can fail even when $A,B$ are tomographically complete, meaning that correlation experiments between $A$ and $B$ fully determine every state on $\H$. Tomographic completeness is equivalent to saying that $A\vee B$, the von Neumann algebra generated by $A$ and $B$, is the full algebra of bounded operators, i.e., $A\vee B=B(\H)$.
Concrete examples are given by the ground states of topologically ordered systems when $A$ consists of two disjoint, infinite cones \cite{naaijkensKosakiLongoIndexClassification2013,fiedlerJonesIndexSecret2017} or in the context of 1+1-dimensional conformal field theories, where $A$ consists of two disjoint intervals of the circle \cite{Kawahigashi2001a}, see Figure~\ref{fig:disjoint-cones}.

Commutativity of $A$ and $B$ can equivalently be formulated as the inclusion $A\subset B'$. Together with tomographic completeness, it implies that $A$ and $B$ are factors ($A\cap A'=\CC$):
\begin{align}\label{eq:no-left-out}
    A'\cap A \subset A' \cap B' = (A\vee B)' = B(\H)' = \CC.
\end{align}
The statement $A'\cap B' =\CC$ means that there are no full quantum degrees of freedom left out in the description, and there are no non-trivial operators in $B'$ that commute with $A$.
It is thus not obvious what impact the remaining operators in $B'\setminus A$ have.
In the mathematics literature, an inclusion $N\subset M$ is called irreducible if the relative commutant $N'\cap M$ is trivial, i.e., $N'\cap M=\CC$. 
Thus, tomographic completeness means that $A\subset B'$ is an irreducible subfactor inclusion.
Subfactor inclusions are an extremely rich field of mathematics, connecting operator algebras with tensor categories, low-dimensional topology, and mathematical physics \cite{jonesIndexSubfactors1983a,jonesPolynomialInvariantKnots1985,Witten1989,turaev_quantum_2010,jonesHeckeAlgebraRepresentations1987,jonesIntroductionSubfactors1997,freedman_topological_2003,kawahigashi_subfactor_2005,longo_index_1989,longo_index_1990,longo_nets_1995,wang_topological_2010,liu_jones-wassermann_2019,bischoffTensorCategoriesEndomorphisms2015,kawahigashi_conformal_2015,hollands_anyonic_2022,kawahigashi_zipper_2026}.

Even if Haag duality fails, some structure remains in the examples mentioned above: 
Namely, there typically exists a \emph{conditional expectation} $E: B' \to A$, i.e., a (normal) unital positive map on $B'$ such that
\begin{align}
	E(abc) = a E(b) c,\quad a,c\in A, b\in B'. 
\end{align}
Equivalently, $E$ is a (normal) unital positive map with range $A$ such that $E^2 = E$.
We say that a state $\omega_{B'}$ on $B'$ is $E$-invariant (or that $E$ is $\omega_{B'}$-preserving) if $\omega_{B'} = \omega_{B'} \circ E$.
This means that $\omega_{B'}$ is completely specified by its marginal $\omega_A$ on $A\subset B'$ and $E$:
\begin{align}
		\omega_{B'}(b) = \omega_{B'}(E(b)) = \omega_A(E(b)).
\end{align}
If the $B'$-marginal $\omega_{B'}$ of a state vector $\Omega\in\H$ is invariant under a conditional expectation $E:B'\to A$, we also say that $E$ is $\Omega$-preserving.
Any state $\omega_A$ on $A$ induces an $E$-invariant state $\omega_{B'}$ on $B'$ by setting $\omega_{B'}  = \omega_A \circ E$ since $E^2 = E$.

Conditional expectations are central objects in the study of operator algebras. 
It is well known that if $\omega$ is a faithful state on $B'$, then Takesaki's theorem \cite[Thm.~IX.4.2]{takesaki2} states that an $\omega$-preserving conditional expectation $B'\to A$ exists if and only if the modular flow of $\omega$ leaves $A$ invariant.
While central to the theory of von Neumann algebras, Takesaki's theorem does not provide a quantum information-theoretic \emph{interpretation} of an $\omega$-preserving conditional expectation. 
This is all the more so in light of the fact that the concept of a conditional probability distribution does not generalize to the quantum setting: There is, in general, no "conditional quantum state". This leads us to ask:
\begin{center}
What are the operational consequences of the existence of an $\omega$-preserving conditional expectation?
\end{center}
The existence of a $\omega$-preserving conditional expectation is a statement about the inclusion $A\subset B'$ alone, independent of the embedding of $B'$ in $B(\H)$. 
However, entanglement enters our setting because we assume that $\omega$ is the $B'$-marginal of a state vector $\Omega \in \H$. 
If $B'$ is a factor not of type $\I$, then this already implies that $B$ and $B'$ are infinitely entangled for every $\Psi \in \H$ \cite{keylInfinitelyEntangledStates2002,verch_distillability_2005,crannStateConvertibilityNeumann2020,vanluijkPureStateEntanglement2025,van_luijk_entanglement_2025}.

We will argue below that (quantum) \emph{steering} provides an operational interpretation of state-preserving conditional expectations.
To explain steering, let us briefly recall the description of measurements in quantum theory.
A measurement with finitely many outcomes is described by a finite \emph{positive operator valued measure (POVM}) $\{a_x\}_{x\in X'}$, which is a collection of positive operators such that $\sum_x a_x = 1$.
We say that a (finite) collection of positive functionals $\{\omega_x\}_{x\in X}$ on a von Neumann algebra $A$ is an \emph{ensemble} if $\sum_x \omega_x$ is a state, called the "average" state of the ensemble.
We often suppress the set $X$, denoting POVMs and ensembles by $\{a_x\}$ and $\{\omega_x\}$, respectively.
If $A$ and $B$ are commuting von Neumann algebras on $\H$ and if $\{b_x\}_{x\in X}$ is a measurement in $B$, then, for any normal state $\omega$ on $B(\H)$, an ensemble $\{\omega_{A,x}\}$ is defined on $A$ by
\begin{equation}\label{eq:steering}
    \omega_{A,x}(a) = \omega(ab_x), \qquad a\in A.
\end{equation}
This ensemble describes the quantum state of the subsystem $A$ conditioned on the outcome of the measurement on $B$.
One says that $B$ \emph{steers} the ensemble $\{\omega_{A,x}\}$ on $A$ using the measurement $\{b_x\}$.
The average state of the ensemble on $A$ is always the marginal $\omega_A:=\omega|_A$ of the global state $\omega$:
\begin{equation}\label{eq:non-signaling}
    \sum_x \omega_{A,x} = \omega_A.
\end{equation}
In particular, the average state is independent of the choice of measurement in $B$.
This is known as the \emph{no-signaling} property.
If $B$ can steer each ensemble on $A$ whose average state is $\omega_A$, we say that \emph{$B$ can steer $A$ relative to $\omega$}.
If $\Omega\in \H$ is a state vector, we will say that steering is possible relative to $\Omega$ if it is possible relative to the pure state $\omega=\ip\Omega{(\placeholder)\Omega}$.

Suppose $B$ steers an ensemble $\{\omega_{A,x}\}$ on $A$ relative to a state $\omega$ using a POVM $\{b_x\} \subset B$.
By considering elements $a\in B'$ instead of $A$ in \eqref{eq:steering}, one obtains an ensemble on $B'$.
The average state of this ensemble is $\omega_{B'}$, the $B'$-marginal of $\omega$, and if we restrict the ensemble to $A\subset B'$, we get $\{\omega_{A,x}\}$.
Thus, for $B$ to be able to steer an ensemble $\{\omega_{A,x}\}$ on $A$ relative to $\omega$ it is necessary that an ensemble extension $\{\omega_{B',x}\}$ on $B'$ with average state $\omega_{B'}$ exists.
Using the Radon-Nikodym theorem for states on von Neumann algebras, we show that this is also sufficient, provided that $\omega$ is a pure state.
Here and throughout the paper, we restrict ourselves to separable Hilbert spaces.

\begin{introlemma}\label{introlem}
    Let $\Omega\in \H$ be state vector and let $\{\omega_{A,x}\}$ be an ensemble on $A$ with $\sum_x\omega_{A,x}=\omega_A$.
    The following are equivalent:
    \begin{enumerate}[(a)]
	\item $B$ can steer $\{\omega_{A,x}\}$ relative to $\Omega$.
	\item \label{introthm:main-it2} 
    The ensemble $\{\omega_{A,x}\}$ can be extended to an ensemble on $B'$ whose average state is $\omega_{B'}$, i.e., there exists an ensemble $\{\omega_{B',x}\}$ with $\omega_{B',x}|_A = \omega_{A,x}$ and $\sum_x \omega_{B',x}= \omega_{B'}$.
    \end{enumerate}
\end{introlemma}

By \cref{introlem}, $B$ can steer $A$ relative to $\Omega$ if and only if the following \emph{ensemble extension property} holds:
Every ensemble on $A$ with average state $\omega_A$ can be extended to an ensemble on $B'$ with average state $\omega_{B'}$.
Note that the ensemble extension property only depends on the inclusion $A\subset B'$ and the states $\omega_{B'}$, but not on the given representation on the Hilbert space $\H$. 

In a finite-dimensional system with $\H = \H_A\ox\H_B\ox\H_E$, $B$ can steer $A$ (we write $X = B(\H_X) \otimes 1_{X^c}$) relative to $\Omega\in \H$ if and only if $A$ and $E$ are uncorrelated, i.e., $\rho_{AE}=\tr_B \kettbra\Omega$ is a product state.
Hence, tomographic completeness, which is equivalent to having a trivial environment, i.e., $\dim\H_E=1$, necessarily guarantees that steering is possible in both directions relative to any pure state.\footnote{Note that this is also true for product states, since pure states only have trivial ensembles, where all elements are proportional to the average state.}
Our main result shows that the general case is very different:
Tomographic completeness does not guarantee steering, and we will see below that it does not imply that one-way steering entails two-way steering.

\begin{introtheorem}\label{introthm:main}
    Let $A,B\subset B(\H)$ be commuting factors, let $\Omega\in\H$ be a state vector.
    The following are equivalent:
    \begin{enumerate}[(a)]
        \item\label{it:intromain1}$B$ can steer $A$ relative to $\Omega$;
        \item\label{it:intromain2} every ensemble on $A$ with average state $\omega_A$ extends to an ensemble on $B'$ with average state $\omega_{B'}$;
        \item\label{it:intromain3} there exists an $\Omega$-preserving normal conditional expectation $E:B'\to A$, i.e., $\omega_{B'}=\omega_A \circ E$.
    \end{enumerate}
	The equivalent conditions imply that every normal state $\psi_A$ on $A$ with $\supp\psi_A\lesssim \supp\omega_A$ has a purification $\Psi\in \H$ that allows $B$ to steer $A$ relative to $\Psi$.
\end{introtheorem}

In fact, we show a slightly stronger version of \cref{introthm:main}: It is enough to require that $B$ can steer all binary ensembles. 
The symbol $\lesssim$ appearing in the last statement of the Theorem denotes the Murray-von Neumann order of projections, essentially asking $\psi_A$ is supported on a smaller subspace than $\omega_A$.

In the case of an irreducible subfactor inclusion $A\subset B'$, there can exist at most one normal conditional expectation $B' \to A$, which is automatically faithful. 
In this case, if $B$ can steer $A$ relative to a single state vector $\Omega$ with faithful $A$-marginal, then every state $\varphi_A$ has a purification $\Phi\in \H$ relative to which $B$ can steer $A$. 
All these purifications are related by unitaries in $B$, since their marginal on $B'$ is fixed to be $\varphi_A \circ E$. 

We emphasize that there exist irreducible subfactor inclusions that do not allow for any normal conditional expectation, even in the case of approximately finite-dimensional (also called hyperfinite) factors.\footnote{For instance, the irreducible subfactor inclusion of the hyperfinite type III$_1$ factor $R_\oo$ into the crossed-product $R_\oo \rtimes \RR$ by the modular flow of a faithful normal state \cite{takesaki2} does not admit a normal conditional expectation.}
Thus, in principle, there are cases where no state on $A$ allows it to be steered. 
It would be interesting to know whether these can be ruled out on general grounds based on physical assumptions, for example, in many-body systems. 

In applications, it is a natural question to ask which states are preserved by the conditional expectation $B'\to A$, should one exist at all. 
For example, in the case of the surface code, in the setup depicted in Fig.~\ref{fig:disjoint-cones}, the ground state is preserved by the conditional expectation from the commutant of the factor associated with the orange region onto the factor associated with the turquoise region \cite{fiedlerJonesIndexSecret2017}.
However, in the case of rational conformal field theories, the restriction of the vacuum state is not preserved, as follows from the observation in the next paragraph together with the Reeh-Schlieder property, i.e., the vacuum is cyclic-separating for every interval \cite{reeh_bemerkungen_1961, borchers_converse_1968, Brunetti1993, Gabbiani1993, Fredenhagen1996}.

Indeed, suppose that the inclusion $A\subset B'\subset B(\H)$ admits a normal conditional expectation $B'\to A$ and let $\Omega\in \H$ be a state vector. 
If $E$ preserves the marginal $\omega_{B'} = \langle \Omega, \placeholder \Omega\rangle|_{B'}$ then 
\begin{align}\label{eq:cyclicity}
    E(b')s_{A'} = s_{A'} b' s_{A'},\quad b'\in B',
\end{align}
where $s_{A'} = [A\Omega] \in A'$ is the cyclic projection of $A$ relative to $\Omega$, which is also the support projection of $\omega_{A'}$.%
\footnote{
To see this, consider $a_1,a_2\in A$ and $b'\in B'$. If $\omega_{B'} = \omega_{B'}\circ E$, we have  for any $a_1,a_2\in A$ and $b'\in B'$ (since $E(a_1^* b' a_2) = a_1^* E(b') a_2$)
\begin{align}
    \langle a_2\Omega, E(b') a_1\Omega\rangle  = \omega_{B'}(a_2^* E(b') a_1) = \omega_{B'}(a_2^* b' a_1) = \langle a_2\Omega, b' a_1 \Omega\rangle.
\end{align}
}
Thus, if $\Omega$ is cyclic for $A$ ($s_{A'} = 1$) we find $E = \id$ and hence $A=B'$.%
\footnote{It is well known that if $N\subset M$ is a finite-index inclusion of type $\II_1$ factors, one can always represent it such that $N$ is in standard form, i.e., admits a cyclic-separating vector $\Omega_N$. This does not contradict our statement: In this representation the trace-preserving conditional expectation $E:M\to N$ does not preserve the marginal of $\Omega_N$ on $M$ unless $N=M$.}

We leave as an open question to find conditions that a) guarantee the existence of a conditional expectation $B'\to A$ for reasonable bipartitions in quantum spin systems, and b) to find conditions that guarantee that the preserved state is the (restriction of) the ground state.

Under the assumption of tomographical completeness, we generalize our characterization of one-way steering to mixed normal states $\omega$ on $B(\H)$.
In addition to a state-preserving conditional expectation, the necessary and sufficient condition is that there is a tensor product decomposition $p\H=\H_0\ox\H_1$, where $p=\supp\omega_B$, such that $B'p\subset B(\H_0)\ox\1$ and $\omega=\omega_0\ox\omega_1$ on $B(p\H)$ with $\omega_0$ a pure state on $\H_0$.
We see that steering relative to mixed states reduces to steering relative to pure states (indeed, in the above factorization, only the pure state $\omega_0$ is seen by the steering).

\subsection{Haag duality, finite-index subfactors and two-way steering}

Steering is a directed property: If $B$ can steer by $A$, it does not automatically follow that $A$ can steer $B$.
Here, we are interested in cases, where two-way steering is possible.
There are two variants: (1) the reference states can be different for the two directions, and (2) for both directions the same reference state should be used.
For tomographically complete setups, we show that the first variant is characterized by $A\subset B'$ (equivalently, $B\subset A'$) being a finite-index subfactor inclusion, while the second variant characterizes Haag duality.
The latter answers an open question in \cite{van_luijk_schmidt_2024} on the relationship between steering and Haag duality.

\begin{introtheorem}\label{introthm:twoway}
Let $A,B\subset B(\H)$ be commuting factors with $A\vee B = B(\H)$. 
Then:
\begin{enumerate}[(1)]
    \item $A$ can steer $B$ relative to a normal state $\omega$ and $B$ can steer $A$ relative to a normal state $\phi$ on $B(\H)$ if and only if the subfactor inclusion $A\subset B'$ has finite index:
    \begin{equation}
        [B':A] <\oo.
    \end{equation}
    If the equivalent properties hold, the states $\omega$ and $\phi$ can be chosen to be pure.
    \item $A$ can steer $B$ and $B$ can steer $A$ relative to the same normal state $\omega$ on $B(\H)$ if and only if $A=B'$ and $\omega$ is pure.
\end{enumerate}
\end{introtheorem}

Here, the index refers to the Jones-Kosaki-Longo index \cite{jonesIndexSubfactors1983a,kosakiExtensionJonesTheory1986,longoIndexSubfactorsStatistics1989a,longoIndexSubfactorsStatistics1990,kosakiRemarkMinimalIndex1992,hiai_minimizing_1988}.
We note that, if one drops tomographical completeness, then two-way steering may be possible even if the index is infinite $[B':A]=\infty$. 
Indeed, take the trivial example $A=B=\CC$ on an infinite-dimensional Hilbert space $\H$ and let $\omega$ be an arbitrary normal state on $B(\H)$.
Then, $A$ can steer $B$ and $B$ can steer $A$ relative to $\omega$, but $A\ne B'$ and $[B':A] = [B(\H):\CC] = \oo$.\footnote{Indeed, if $A=\CC$, then every ensemble $\{\omega_{A,x}\}$ is of the form $\omega_{A,x}= p_x\omega_A$ for a probability distribution $\{p_x\}$ and can be steered using the POVM $\{p_x\} \subset B=\CC$.} Item (1) of \cref{introthm:twoway} is proved as \cref{prop:index} and item (2) as \cref{prop:asymmetry}.

\subsection{Lifting ensembles under general quantum channels}

The lifting property for subalgebra ensembles with a fixed average state considered above can be regarded as a special case of a more general property.
Consider two pairs $(N,\phi)$ and $(M,\omega)$ of von Neumann algebras equipped with normal states.
Let 
\begin{equation}
    \alpha: (N,\phi)\to (M,\omega)
\end{equation}
be a state-preserving quantum channel, i.e., $\alpha:N\to M$ is a normal unital completely-positive map such that $\phi = \omega\circ\alpha$.
Does every functional dominated by $\phi$ lift to a functional dominated by $\omega$ under the map $\alpha$?
In the case that $\alpha$ is an inclusion $A \hookrightarrow B'$, this question reduces to the ensemble extension property considered above.

\begin{introtheorem}\label{introthm:lifting}
    Let $\alpha : (N,\phi) \to (M,\omega)$ be a state-preserving normal unital cp map.
    Suppose $\phi$ is faithful.
    Then the following are equivalent:
    \begin{enumerate}[(a)]
        \item\label{it:intro-lifting1} 
         For every ensemble $\{\phi_x\}$ on $N$ with average state $\sum_x\phi_x=\phi$, there is an ensemble $\{\omega_x\}$ on $M$ with average state $\sum_x\omega_x=\omega$ such that $\phi_x= \omega_x\circ\alpha$.
        \item\label{it:intro-lifting2} For every $\phi_1\in N_*$ with $0\le\phi_1\le \phi$, there exists $\omega_1\in M_*$ with $0\le \omega_1\le \omega$ and $\phi_1=\omega_1\circ\alpha$.
        \item\label{it:intro-lifting4}  $\alpha$ has a unital completely positive left inverse, i.e., there exists a normal unital completely positive map $\beta:M\to N$ with $\beta\circ\alpha = \id$.
        \item\label{it:intro-lifting3}
        $\hat\alpha = P_{\supp \omega}\circ \alpha$ is a *-isomorphism onto a subalgebra $L\subset M_{\supp\omega}$ and this subalgebra admits a $\omega$-preserving conditional expectation $E:M_{\supp\omega} \to L$. Here, $P_{\supp \omega}: M\to M_{\supp\omega}$ is the cut-down onto the support corner.
    \end{enumerate}
\end{introtheorem}

Let us comment on the easy implications.
\ref{it:intro-lifting1} implies \ref{it:intro-lifting2} by applying the condition to binary ensembles $\{\phi_1,\phi-\phi_1\}$.
\ref{it:intro-lifting3} implies \ref{it:intro-lifting4} via $\beta = \hat\alpha^{-1}\circ E \circ P_{\supp \omega} + (1-\supp\phi)\omega$
.
\ref{it:intro-lifting3} implies \ref{it:intro-lifting1} by defining the ensemble lifts as $\omega_x = \phi_x\circ \beta$.
Using multiplicative domain properties and the theory of Petz dual maps, we will then show the missing implications \ref{it:intro-lifting2} $\Rightarrow$ \ref{it:intro-lifting4} $\&$ \ref{it:intro-lifting3}.

For the lifting property in the first item to make sense, it is not necessary that $\alpha$ be completely positive; positivity alone suffices.
Motivated by this, we prove the theorem more generally in the category of JBW*-algebras and normal unital positive maps.
Jordan algebras have recently been found to be the natural framework to formulate and prove certain results in quantum information theory that do not require complete positivity \cite{frenkel_integral_2023,muller-hermes_monotonicity_2017,van_luijk_sufficiency_2026,sonoda_hypothesis_2025,galke_sufficiency_2024}. We hope that our more general proof will be of independent interest also from that point of view.

We show that, if one applies \cref{introthm:lifting} to the inclusion map $N\hookrightarrow M$ of a subfactor, one can drop the support projections in the statement of the theorem.
As a consequence, one obtains the following state-based/Schrödinger-picture characterization of state-preserving conditional expectations (as opposed to Takesaki's observable-based/Heisenberg-picture theorem):

\begin{introproposition}\label{introprop:qi-takesaki}
    Let $N\subset M$ be von Neumann algebras. Let $\omega$ be a normal state on $M$ and let $\phi=\omega|_N$ be its restriction to $N$.
    Suppose $\phi$ has central support $1$.
    The following are equivalent:
    \begin{enumerate}[(a)]
        \item For every $\phi_1\in N_*$ with $0\le\phi_1\le \phi$, there exists $\omega_1\in M_*$ with $0\le \omega_1\le \omega$ and $\phi_1=\omega_1|_N$.
        \item There exists a normal conditional expectation $E:M\to N$ with $\omega = \phi\circ E$.
    \end{enumerate}
\end{introproposition}

Applying this corollary to the subfactor $A\subset B'$ discussed in the context of quantum steering, and using \cref{introlem}, proves our main result \cref{introthm:main}.

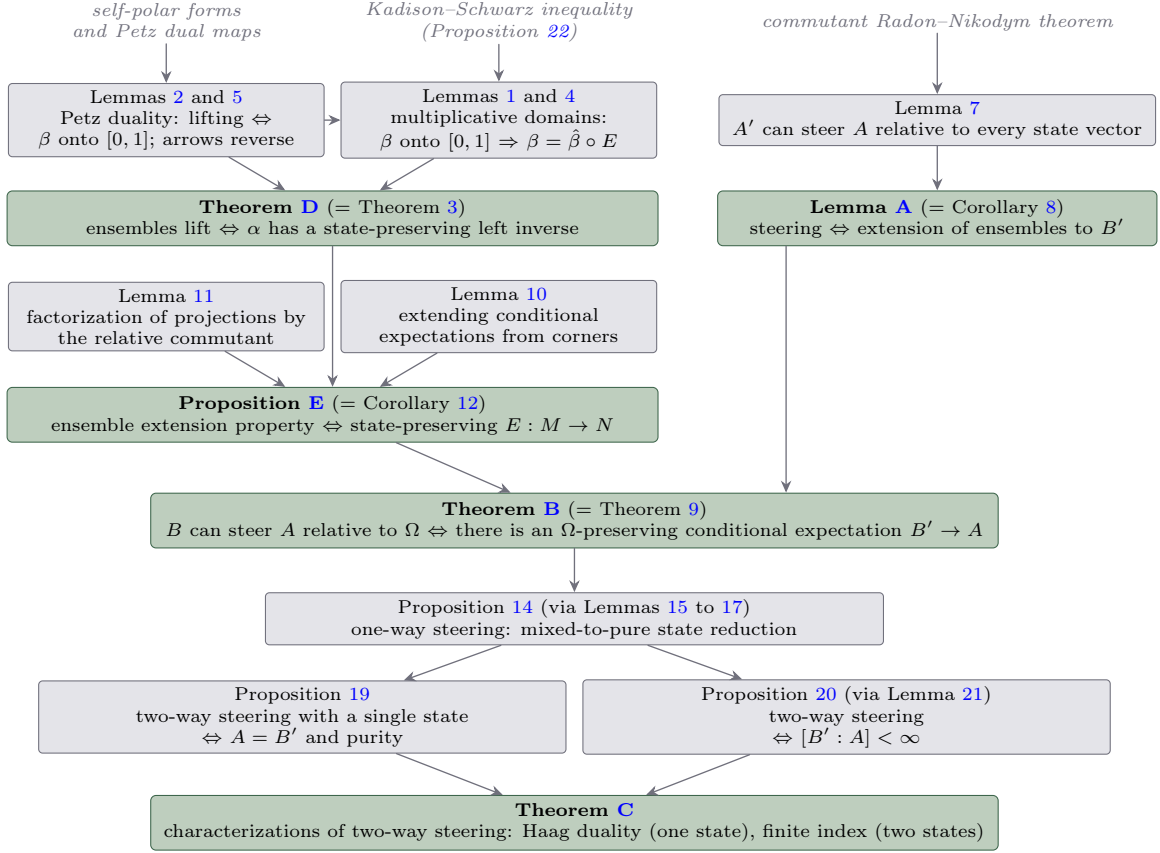
\begin{figure}[!t]
  \centering
  \begin{tikzpicture}[
      x=1mm,y=1mm,line join=round,line cap=round,
      >={Stealth[length=1.6mm,width=1.4mm]},
      ing/.style={font=\scriptsize\itshape,text=Gapdark,align=flush center,
                  inner sep=1pt,text width=#1},
      ing/.default=40mm,
      aux/.style={draw=Gapdark,fill=Gapcol,rounded corners=1.5pt,inner sep=2.5pt,
                  align=flush center,font=\scriptsize,text width=#1},
      aux/.default=40mm,
      res/.style={draw=Adark,fill=Acol,rounded corners=2pt,inner sep=3pt,
                  align=flush center,font=\scriptsize,text width=#1},
      res/.default=84mm,
      ar/.style={->,draw=Gapdark,line width=.5pt},
    ]
    \node[ing]        (ing1) at ( 66,110) {Kadison--Schwarz inequality\\
                                          (\cref{prop:kadison-schwarz})};
    \node[ing]        (ing2) at ( 22,110) {self-polar forms\\ and Petz dual maps};
    \node[ing=56mm]   (ing3) at (124,110) {commutant Radon--Nikodym theorem};
    \node[aux]        (npd)  at ( 22, 97) {\cref{lem:order-interval,lem:functor}\\
                                           Petz duality: lifting $\Leftrightarrow$\\
                                           $\beta$ onto $[0,1]$; arrows reverse};
    \node[aux]        (nmd)  at ( 66, 97) {\cref{lem:multiplicative-dom,lem:surjective}\\
                                           multiplicative domains:\\
                                           $\beta$ onto $[0,1]$ $\Rightarrow$ $\beta=\hat\beta\circ E$};
    \node[res]        (nlift) at ( 44, 84) {\textbf{\Cref{introthm:lifting}} ($=$ \cref{thm:lifting})\\
                                           ensembles lift $\Leftrightarrow$ $\alpha$ has a state-preserving left inverse};
    \node[aux=56mm]   (nst)  at (124, 97) {\cref{lem:hd-implies-steering}\\
                                           $A'$ can steer $A$ relative to every state vector};
    \node[res=56mm]   (nens) at (124, 84) {\textbf{\Cref{introlem}} ($=$ \cref{cor:steer-iff-extend})\\
                                           steering $\Leftrightarrow$ extension of ensembles to $B'$};
    \node[aux]        (ncut) at ( 22, 71) {\cref{lem:cut-down-iso}\\
                                           factorization of projections by the relative commutant};
    \node[aux]        (next) at ( 66, 71) {\cref{lem:extending-CEs}\\
                                           extending conditional\\
                                           expectations from corners};
    \node[res]        (ntak) at ( 44, 58) {\textbf{\Cref{introprop:qi-takesaki}} ($=$ \cref{prop:qi-takesaki})\\
                                           ensemble extension property $\Leftrightarrow$ state-preserving $E:M\to N$};
    \node[res=110mm]  (nmain) at ( 76, 44) {\textbf{\Cref{introthm:main}} ($=$ \cref{thm:main})\\
                                           $B$ can steer $A$ relative to $\Omega$ $\Leftrightarrow$ there is an $\Omega$-preserving conditional expectation $B'\to A$};
    \node[aux=80mm]   (nmix) at ( 76, 31) {\cref{prop:mixed-one-way} (via \cref{lem:irreducible-up-to-type-I-CE,lem:uncorrelated-purification,lem:steering-support-B})\\
                                           one-way steering: mixed-to-pure state reduction};
    \node[aux=68mm]   (nasy) at ( 40, 18) {\cref{prop:asymmetry}\\
                                          two-way steering with a single state  \\
                                          $\Leftrightarrow$ $A=B'$ and purity};
    \node[aux=68mm]   (nidx) at (112, 18) {\cref{prop:index} (via \cref{lem:purification})\\
                                          two-way steering \\
                                          $\Leftrightarrow$ $[B':A]<\oo$};
    \node[res=110mm]  (ntwo) at ( 76,  4) {\textbf{\Cref{introthm:twoway}}\\
                                           characterizations of two-way steering: Haag duality (one state), finite index (two states)};
    \draw[ar] (ing1) -- (nmd);
    \draw[ar] (ing2) -- (npd);
    \draw[ar] (ing3) -- (nst);
    \draw[ar] (npd)  -- (nmd);
    \draw[ar] (nmd)  -- (nlift);
    \draw[ar] (npd)  -- (nlift);
    \draw[ar] (nst)  -- (nens);
    \draw[ar] (nlift) -- (ntak);
    \draw[ar] (ncut) -- (ntak);
    \draw[ar] (next) -- (ntak);
    \draw[ar] (ntak) -- (nmain);
    \draw[ar] ([xshift=-20mm]nens.south) -- ([xshift=28mm]nmain.north);
    \draw[ar] (nmain) -- (nmix);
    \draw[ar] (nmix) -- (nasy);
    \draw[ar] (nmix) -- (nidx);
    \draw[ar] (nasy) -- (ntwo);
    \draw[ar] (nidx) -- (ntwo);
  \end{tikzpicture}
  \caption{
    The figure provides an overview of the logical dependencies and the proof structure of the results stated in the introduction.
    The left-hand column is the general theory of \cref{sec:positive-maps}, which is carried out for normal unital \emph{positive} maps between JBW*-algebras. the right-hand column provides the basic observations on steering.
    \Cref{introthm:main} combines the two:
    \cref{introlem} turns steering into the ensemble extension property, \cref{introprop:qi-takesaki} turns the latter into a state-preserving conditional expectation.
    Both items of \cref{introthm:twoway} come out of \cref{introthm:main} via the reduction of mixed states to pure ones.}
  \label{fig:roadmap}
\end{figure}

The logical dependencies and the organization of the proofs of the stated results are illustrated in Figure \cref{fig:roadmap}.

\null

\paragraph{Acknowledgements.} H.W. would like to thank Pieter Naaijkens and Daniel Wallick for discussions on state-preserving conditional expectations in many-body systems.
We acknowledge the help of Claude Code to derive the Kadison-Schwarz inequality for general JB-algebras as well as for proof checking.

\paragraph{Funding.} Research at Perimeter Institute and the University of Waterloo is supported in part by the Government of Canada through the Department of Innovation, Science and Economic Development and by the Province of Ontario through the Ministry of Colleges and Universities.

Funded by the European Union. Views and opinions expressed are however those of the authors only and do not necessarily reflect those of the European Union or the European Research Council Executive Agency. Neither the European Union nor the granting authority can be held responsible for them.
This work is supported by an 
ERC grant (ERC StG LargEnt, 101219447, \href{https://doi.org/10.3030/101219447}{DOI:10.3030/101219447}).

Funded by zukunft.niedersachsen, the joint science funding program of the
Lower Saxony Ministry of Science and Culture and the Volkswagen
Foundation.

\paragraph{Contributions and assistance by large-language models (LLMs).} 
The initial idea for this paper arose around discussions between LvL, AS and HW. LvL conjectured the equivalence of the ensemble extension property with the existence of a state-preserving CE and realized the connection of ensemble extensions to steering. 
TJO contributed a first handwritten (how quaint!) argument for a finite-dimensional case. 
This was then extended by LvL, AS and HW to a class of von Neumann algebras.
AM found a much cleaner, multiplicative domain-based argument, which extended the results to arbitrary von Neumann algebras.
LvL observed that the proof applied almost verbatim to the case of JBW* algebras. 
TJO utilized Claude Fable 5 and GPT 5.5 to formalize the arguments in the paper, which exposed a small gap in the proof, which was closed by LvL, AS and HW. 
HW worked with Claude Opus 4.8 to prove the Kadison-Schwarz inequality for JB algebras, which we assumed to be in the literature but could not find. 
The steering-related statements were shown by LvL, AS and HW, with LLMs used for proof checking and to realize that Lemma 21 is true.
LLMs were occasionally used by LvL, AS and HW to review various topics in subfactor theory. 
All references in the bibliography were inserted manually.
LLMs were also used for grammar checking and creating an initial version of \cref{fig:roadmap}.

\section{Lifting ensembles under positive maps}
\label{sec:positive-maps}

We begin by proving a generalized version of \cref{introthm:lifting} which holds for maps that are merely positive instead of being completely positive.
Instead of assuming von Neumann algebras, we allow $N$ and $M$ to be JBW*-algebras.
In the setting of a completely positive map between von Neumann algebras, our proof immediately spits out the completely positive version of the theorem.

\subsection{JBW*-algebras and normal unital positive maps}

We choose the complex setting of JBW*-algebras instead of the real setting of JBW-algebras to smoothen the transition to the setting of von Neumann algebras.
We refer to \cite{hanche-olsen_jordan_1984} for the definition and basic properties of JBW*-algebras.
However, let us briefly recall a few key facts.

A JBW*-algebra is in particular a unital \emph{Jordan *-algebra}, i.e., a complex vector space $M$ closed under a conjugate-linear involution $*$ and carrying a commutative, but possibly non-associative, bilinear product $(a,b)\mapsto a \circ b$ fulfilling $(a\circ b)^* = a^*\circ b^*$ and the \emph{Jordan identity} $a\circ(b\circ a^2) = (a\circ b) \circ a^2$ on self-adjoint elements $a,b\in M$.
Here, we already took the liberty to write $a^2 = a\circ a$. A useful algebraic tool is the 
\emph{Jordan triple-product} $\{abc\} = (a\circ b)\circ c +  a\circ(b\circ c)  - (a\circ c)\circ b$.

In addition, a JBW*-algebra has analytical properties: It comes with a norm, turning it into the Jordan analog of a C*-algebra (called a JB*-algebra), and with a (necessarily unique) Banach space predual, making it the Jordan version of a von Neumann algebra.
To avoid technicalities, we will only consider JBW*-algebras (and von Neumann algebras) with separable predual.
The weak* topology induced by the predual $M_*$ of a JBW*-algebra $M$ is called the ultraweak topology.
For every self-adjoint element $a$ of a JBW*-algebra $M$, the JBW*-subalgebra $W^*(a)\subset M$ that is generated by $a$ is an abelian von Neumann algebra.
A self-adjoint element $a \in M$ is positive if it is positive as an element of $W^*(a)$, or equivalently, it is positive if it is the square $a=b^2$ of some other self-adjoint element $b\in M$. 

Every von Neumann algebra $M$ is a JBW*-algebra with product $a\circ b = \frac{1}{2}(ab +ba)$.
A weakly closed, *-invariant unital subspace $N\subset M$ of a von Neumann algebra is a JBW*-algebra (in fact, a so-called JW*-algebra) if it is closed under the Jordan product.
For instance, if $\vartheta$ is an anti-automorphism on a von Neumann algebra $M$, then its fixed point space $N = \{a\in M : \vartheta(a)=a\}$ is a JBW*-algebra.
In a von Neumann algebra, the Jordan triple is simply the symmetrized product $\{abc\} = \frac12(abc+cba)$.
In particular, $\{aba\}$ is just $aba$.
We refer to \cite{hanche-olsen_jordan_1984} for more examples.

The self-adjoint part of JBW*-algebra is a JBW-algebra and, by Wright's Theorem  \cite{wright_jordan_1977}, every JBW*-algebras arises by complexifying a JBW-algebra, see \cite[Sec.~3.8]{hanche-olsen_jordan_1984}.
While most of the literature is written for JBW-algebras, the results that we reference follow by elementary complexification from the stated results for JBW-algebras.
The one exception is the Kadison-Schwarz inequality for unital positive maps between JBW-algebras, proven in the appendix.

We consider normal unital positive maps $\alpha: N\to M$ between JBW*-algebras $N$ and $M$.
Here, normality means that $\alpha$ is continuous for the respective ultraweak topologies.
In particular, this includes normal J*-homomorphisms.
Here, a normal J*-homomorphism (resp.\ J*-isomorphism) between JBW*-algebras is a normal *-preserving Jordan homomorphism (resp.\ isomorphism) between Jordan *-algebras.
\footnote{In the case of J*-isomorphisms, normality is automatic. 
Indeed, as for *-isomorphisms between von Neumann algebras, this follows from the uniqueness of the predual \cite[Thm.~4.16]{hanche-olsen_jordan_1984}. }

The theory of multiplicative domains for completely positive maps rests on the (Kadison-)Schwarz inequality, which fails for general positive maps.
There is, however, a generalization, called the \emph{Jordan-Schwarz inequality}, which is valid for normal unital positive maps between JBW*-algebras and which allows for a generalization of multiplicative domains to the setting of normal unital positive maps between JBW*-algebras:

\begin{lemma}\label{lem:multiplicative-dom}
    Let $\alpha: N \to M$ be a normal unital positive map between JBW*-algebras.
    Then
    \begin{equation}\label{eq:JS-inequality}
        \alpha(a)\circ \alpha(a^*) \le \alpha(a\circ a^*), \qquad a\in N.
    \end{equation}
    If $\mult(\alpha)$ is the set of elements $a\in N$ for which \eqref{eq:JS-inequality} holds with equality, then:
    \begin{enumerate}
        \item\label{it:mult1} $a\in \mult(\alpha)$ if and only if $\alpha(a\circ b)=\alpha(a)\circ \alpha(b)$ for all $b\in N$;
        \item\label{it:mult2} $\mult(\alpha)$ is a JBW*-subalgebra of $N$.
    \end{enumerate}
\end{lemma} 

\begin{proof}
We follow \cite[Prop.~2.1.5, Prop.~2.1.7]{stormer_positive_2013}, where the corresponding statements are proved for C*-algebras. 
First, consider a self-adjoint element $a\in N$. Then by the Kadison-Schwarz inequality for unital positive maps between JB-algebras shown in \cref{prop:kadison-schwarz} of the appendix, we have $\alpha(a^2) \geq \alpha(a)^2$.  Now consider a general element $a\in N$ and set $b = a+a^*$ and $c= i(a-a^*)$. Then
\begin{align}
    4\alpha( a\circ a^*) = \alpha(b^2) + \alpha(c^2) \geq \alpha(b)^2 + \alpha(c)^2 = 4 \alpha(a)\circ \alpha(a^*),
\end{align}
showing \cref{eq:JS-inequality}. 

Item \ref{it:mult1}: 
First note that $\mult(\alpha)$ is closed under adjoints. Now let $a \in \mult(\alpha)$, $b\in N$ and $t\in \RR$ be self-adjoint and apply the Jordan-Schwarz inequality to the element $ta+b$  to get 
\begin{align}
    2t\,\alpha(a\circ b) + \alpha(b^2) \geq 2t\, \alpha(a)\circ \alpha(b) + \alpha(b)^2. 
\end{align}
Dividing by $2t$ and taking the limits $t\to \pm \infty$ we find
\begin{align}
    \alpha(a\circ b) \geq \alpha(a)\circ \alpha (b) \geq \alpha(a\circ b).  
\end{align}
This shows \cref{it:mult1} for self-adjoint elements, but the argument extends to general $b\in N$ by linearity:
Let $a= b + ic \in \mult(\alpha)$ with $b,c\in N_{sa}$. Since $\alpha(a \circ a^*) = \alpha(a)\circ \alpha(a^*)$ it follows that
\begin{align}
   \left( \alpha(b^2) -\alpha(b)^2\right) + \left( \alpha(c^2)  - \alpha(c)^2\right)  = 0. 
\end{align}
Since both summands are positive by \cref{eq:JS-inequality}, we find that $b,c\in \mult(\alpha)$. 
By linearity, this shows \cref{it:mult1} for general arguments.

Item \ref{it:mult2}:
Since $\alpha$ is a normal positive map and since multiplication by a fixed element is ultraweakly continuous, $\mult(\alpha)\subset N$ is a unital, ultraweakly closed, *-invariant subspace.
We have to show that it is closed under the Jordan product.
By polarization, it suffices to show that $\mult(\alpha)$ contains the square of each hermitian element.
By \eqref{eq:JS-inequality}, this amounts to showing $\alpha(a^2)^2=\alpha(a^4)$ for  $a=a^*\in \mult(\alpha)$.
Iterating the first item, we have
\begin{equation}
    \alpha(a^4) = \alpha(a \circ a^3) =\alpha(a)\circ\alpha(a^3) = \alpha(a) \circ (\alpha(a) \circ \alpha(a^2)) = \alpha(a) \circ (\alpha(a)\circ \alpha(a)^2)
    = \alpha(a)^4.
\end{equation}
Thus, by \eqref{eq:JS-inequality}, we have
\begin{equation}
    \alpha(a^4) \ge \alpha(a^2)^2 \ge \alpha(a)^4 = \alpha(a^4).
\end{equation}
Therefore, equality holds everywhere, so that $\alpha(a^2)^2=\alpha(a^4)$. This proves $a^2\in\mult(\alpha)$.
\end{proof}

We refer to the JBW*-subalgebra $\mult(\alpha)\subset N$ as the \emph{multiplicative domain} of $\alpha$.
In the literature, this name is usually reserved for completely positive maps.\footnote{For instance, Størmer refers to $\mult(\alpha)$ as the "definite set" of $\alpha$ \cite{stormer_multiplicative_2007,stormer_positive_2013}.}
We use for general positive maps since, if $\alpha$ is a completely positive between von Neumann algebras, $\mult(\alpha)$ is exactly the multiplicative domain in the usual sense (in particular, it is a von Neumann subalgebra).

Suppose $N\subset M$ are JBW*-algebras.
A (normal) \emph{conditional expectation} of $M$ onto $N$ is a normal unital positive map $E: M\to N$ with $E|_N =\id$. 
In the literature, the term conditional expectation is sometimes reserved for associative operator algebras. 
Using the term in the broader context of JBW*-algebras is justified by Tomiyama's theorem \cite{tomiyama_projection_1957}, which implies that a conditional expectation in our sense is also a conditional expectation in the usual sense whenever $M$ and $N$ are von Neumann algebras.

Next, we need the concept of \emph{self-polar} forms, introduced by Woronowicz in \cite{woronowicz_selfpolar_1974}, and studied in the context of JBW*-algebras in \cite{haagerup_tomita-takesaki_1984}.
For a normal positive sesquilinear form  $s$ on a JBW*-algebra $M$, we denote by $\Gamma_s$ the anti-linear positive map $M \to M_*$ given by $\Gamma_s(x) = s(x,\placeholder)$.
A (normal) self-polar form on a JBW*-algebra is a normal positive sesquilinear form $s:M\times M\to \RR$ with the following properties:
\begin{enumerate}[(i)]
    \item $s(x,y)\ge0$ for all $0\le x,y\in M$,
    \item $\Gamma_s([0,1]) = [0,\Gamma_s(1)]$.
\end{enumerate}
If $s$ is a self-polar form such that $s(1,1)=1$, then $\omega = \Gamma_s(1)$ defines a normal state on $M$.
Conversely, it is shown in \cite{haagerup_tomita-takesaki_1984} that for every normal state $\omega$, there is a unique self-polar form $s$ with $\Gamma_s(1)=\omega$, where the uniqueness is a consequence of Woronowicz's maximum principle \cite{woronowicz_selfpolar_1974}.

We denote this self-polar form by $s_\omega$ and set $\Gamma_\omega := \Gamma_{s_\omega}$.
The restriction of $s_\omega$ to $M_{\supp\omega}$ is non-degenerate.
Therefore, $\Gamma_\omega$ restricts to an anti-linear order isomorphism between $M_{\supp \omega}$ and $\lin\,[0,\omega]$.

In particular, if $\omega$ is faithful, then $\Gamma_\omega$ is an order isomorphism of $M$ and $\lin\,[0,\omega]$.
When $M$ is a von Neumann algebra in standard representation, the self-polar form can be expressed using the GNS representation with respect to $\omega$ as
\begin{equation}
    s_\omega(a,b) = \ip{\xi_\omega}{b J_\omega a\xi_\omega}, \qquad a,b\in M.
\end{equation}
Using Haagerup $L^p$-spaces, it may be written as $s_\omega(a,b) = \tr( a^* \omega^{1/2}b \omega^{1/2})$, and the map $\Gamma_\omega$ may be written as $\Gamma_\omega = \omega^{1/2}(\placeholder)\omega^{1/2}$, where we identify $L^1(M)$ with $M_*$. 

We can now define Petz dual maps \cite{accardi_conditional_1982,petzDUALNEUMANNALGEBRAS1984,petzSUFFICIENCYCHANNELSNEUMANN1988,ohya_quantum_1993,jencovaSufficiencyQuantumStatistical2006} for maps between JBW*-algebras.
Let $\alpha :(N,\phi)\to (M,\omega)$ be a state-preserving normal unital positive map between JBW*-algebras. 
Suppose $\phi$ is faithful ($\omega$ need not be faithful).
Then the dual map of $\alpha$ is the state-preserving normal unital positive map 
$\beta:(M,\omega)\to (N,\phi$) defined by
\begin{equation}
    \beta = (\Gamma_\phi)^{-1} \circ \alpha_* \circ \Gamma_\omega.
\end{equation}
This is well-defined since the pre-adjoint map $\alpha_*$ takes $[0,\omega] = \Gamma_\omega([0,1]_M)$ into $[0,\phi]$.
The definition is best understood through the following commutative diagram:
\begin{equation}
\begin{tikzcd}
    M \arrow{r}{\beta} \arrow{d}{\Gamma_\omega} &N \\[12pt]
    \lin\, [0,\omega] \arrow{r}{\alpha_*} & \lin\, [0,\phi] \arrow{u}{(\Gamma_{\phi})^{-1}}
\end{tikzcd}
\end{equation}
The equation defining $\beta$ may also be written as
\begin{equation}
    s_\omega(x,\alpha(y)) = s_{\phi}(\beta(x),y), \qquad x\in M,\, y\in N.
\end{equation}
We note that if $\omega$ is faithful, then $\beta$ is a faithful map.
Indeed, let $0\le x \in M$ with $\beta(x)=0$. Then $\alpha_*(\Gamma_\omega(x)) = \Gamma_\phi(\beta(x)) = 0$. Evaluating this functional at $1\in N$ and using that $\alpha$ is unital gives
\begin{equation}
    \omega(x) = \Gamma_\omega(x)(1) = \Gamma_\omega(x)(\alpha(1)) = \alpha_*(\Gamma_\omega(x))(1) = 0,
\end{equation}
so that $x=0$ by faithfulness of $\omega$.
Next, we mention functorial properties of the Petz dual.

\begin{lemma}\label{lem:functor}
    Let $(L,\chi), (N,\phi), (M,\omega)$ be pairs of JBW*-algebras with faithful normal states. 
    \begin{enumerate}
        \item\label{it:functor0} Suppose $M,N$ are von Neumann algebras. Let $\alpha:(N,\phi)\to (M,\omega)$ be a state-preserving normal unital positive map, and let $\beta : (M,\omega)\to(N,\phi)$ be its Petz dual.
        Then $\alpha$ is completely positive if and only if $\beta$ is.
        \item\label{it:functor1} Let $\alpha:(N,\phi)\to (M,\omega)$ be a state-preserving normal unital positive map, and let $\beta : (M,\omega)\to(N,\phi)$ be its Petz dual
        Then the Petz dual of $\beta$ is $\alpha$.
        \item\label{it:functor2} Let $\alpha_1 : (N,\phi) \to (L,\chi)$, $\alpha_2 : (L,\chi) \to (M,\omega)$ be state-preserving normal unital positive maps. Let $\beta_1,\beta_2$ be their Petz dual maps. 
        Then the Petz dual map of $\alpha = \alpha_2\circ \alpha_1$ is $\beta = \beta_1 \circ \beta_2$. 
        \item\label{it:functor3} Let $\alpha : (N,\phi) \to (M,\omega)$ be a state-preserving J*-isomorphism. 
        Then its Petz dual is its inverse $\alpha^{-1}$.
        \item\label{it:functor4}
        If $N\subset M$ and if $E:M\to N$ is a conditional expectation with $\omega= \phi\circ E$, then the  Petz dual of $E$ is the embedding $\iota:N\hookrightarrow M$.
    \end{enumerate}
\end{lemma}
\begin{proof}
    Item \ref{it:functor0} follows from the definition of the Petz dual and the fact that a normal linear map $\alpha:M\to N$ between von Neumann algebras is completely positive if and only if its predual $\alpha_*:N_*\to M_*$ is.
    Items \ref{it:functor1} and \ref{it:functor2} follow directly from the definition of the Petz dual.
    Item \ref{it:functor3}: If $\alpha$ is a J*-isomorphism, $s := s_\omega(\alpha \times \alpha)$ is a self-polar form with $\Gamma_s(1)=\phi$. 
    Woronowicz's maximum principle ensures $s_\omega \circ (\alpha\times \alpha) = s_\phi$ \cite{woronowicz_selfpolar_1974}.
    This is equivalent to $\Gamma_\phi = \alpha_*\circ \Gamma_\omega\circ \alpha$. Hence, $\alpha^{-1} = (\Gamma_\phi)^{-1}\circ \alpha_*\circ\Gamma_\omega$ and the right-hand side is precisely the Petz dual of $\alpha$.
    Item \ref{it:functor4}: 
    Note that the assumption of $\omega=\phi\circ E$ implies $\omega|_N =\phi$.
    As shown \cite[Proof of Thm.~4.2, (i)\,$\Rightarrow$\,(ii)]{haagerup_positive_1995}, one has $s_\phi \circ (E\times E) = s_\omega$.
    Thus, we have 
    \begin{equation}
        \Gamma_\omega = E_* \circ \Gamma_\phi \circ E
    \end{equation}
    By composing with $\iota$ from the right and using $E\circ \iota = \id$, we get $\Gamma_\omega \circ \iota = E_* \circ \Gamma_\phi$.
    Therefore, 
    \begin{equation}
    \text{(dual map of $E$)} = (\Gamma_\omega)^{-1} \circ E_* \circ \Gamma_\phi 
    = (\Gamma_\omega)^{-1} \circ \Gamma_\omega \circ \iota = \iota.
    \end{equation}
\end{proof}

The third item in \cref{lem:functor} can be summarized by saying that passing to Petz duals preserves commutative diagrams (but reverses arrows):

\begin{equation}
    \begin{tikzcd}
        (N,\phi) \arrow{rr}{\alpha}\arrow{dr}{\alpha_1} && (M,\omega) \\
        & (L,\chi) \arrow{ur}{\alpha_2}
    \end{tikzcd}
    \ \ \quad\longleftrightarrow\quad\ \ 
    \begin{tikzcd}[arrows=<-]
        (N,\phi) \arrow{rr}{\beta}\arrow{dr}{\beta_1} && (M,\omega) \\
        & (L,\chi) \arrow{ur}{\beta_2}
    \end{tikzcd}
\end{equation}

To treat cases where the states are not faithful, we need to consider corners of JBW*-algebras.
If $e\in M$ is a projection, the corner $eMe$ is the linear hull of elements $0\le a \le e$.
Equivalently, it is set of elements $\{eae\}\equiv eae$, $a\in M$ (note that $\equiv$ is an actual equality if $M$ is a subalgebra of a von Neumann algebra).
The corner is itself a JBW*-algebra with the operations inherited from $M$.

\subsection{Lifting ensembles}

An ensemble on a JBW*-algebra $M$ is a collection $\{\omega_x\}_{x\in X}$ of normal positive linear functional $0\le \omega_x\in M_*$ such that $\sum_x \omega_x(1) = 1$.
In this case, $\sum_x \omega_x$ is a normal state on $M$, called the average state of the ensemble.
We often suppress the index set $X$, which will always be finite or countably infinite.

If $e$ is a projection in a JBW*-algebra $M$, we denote by $P_e$ the cut-down map $x \mapsto exe$ onto the corner $eMe\subset M$. 
If $e = \supp \omega$ is the support projection of a normal state $\omega$, we denote the corresponding corner by $M_{\supp\omega}$.

\begin{theorem}\label{thm:lifting}
    Let $\omega$ and $\phi$ be normal states on JBW*-algebras $M$ and $N$, respectively.
    Let $\alpha : (N,\phi)\to (M,\omega)$ be a state-preserving normal unital positive map.
    The following are equivalent:
    \begin{enumerate}[(a)]
        \item\label{it:lifting1}
        For every ensemble $\{\phi_x\}$ on $N$ with average state $\sum_x\phi_x=\phi$, there is an ensemble $\{\omega_x\}$ on $M$ with average state $\sum_x\omega_x=\omega$ such that $\phi_x= \omega_x\circ\alpha$.
        \item\label{it:lifting2}
        for every $\phi_1\in N_*$ with $0\le \phi_1\le \phi$, there exists $\omega_1\in M_*$ with $0\le \omega_1\le \omega$ and $\phi_1=\omega_1\circ\alpha$.
        \item\label{it:lifting3}
        there exists a state-preserving normal unital positive map $\beta : (M,\omega)\to (N,\phi)$ such that $P_{\supp\phi}\circ \beta\circ \alpha |_{N_{\supp\phi}} = \id_{N_{\supp\phi}}$;
        \item\label{it:lifting4}
        $\hat\alpha = P_{\supp\omega} \circ\alpha|_{N_{\supp \phi}}$ is a J*-isomorphism of $N_{\supp\phi}$ onto a JBW*-subalgebra $L\subset M_{\supp\omega}$, which admits a $\omega$-preserving normal conditional expectation $E : M_{\supp\omega}\to L$.
    \end{enumerate}
    If $M$ and $N$ are von Neumann algebras and if $\alpha$ is completely positive, then $\beta$ can be chosen to be completely positive, $L$ is a von Neumann algebra and $\hat\alpha$ is a *-isomorphism.
\end{theorem}

We begin by characterizing normal unital positive maps that have right inverses.
We will then use Petz duality to obtain a characterization of maps with left inverses.
\begin{lemma}\label{lem:surjective}
    Let $\beta : M \to N$ be a faithful normal unital positive map between JBW*-algebras.
    The following are equivalent:
    \begin{enumerate}[(a)]
        \item\label{it:surjective1} $\beta$ maps $[0,1]_{M}$ onto $[0,1]_{N}$;
        \item\label{it:surjective2}
        The restriction $\hat\beta=\beta|_{\mult(\beta)}$ to the multiplicative domain $\mult(\beta)$ is a J*-isomorphism onto $N$;
        \item\label{it:surjective3}
        $\beta$ admits a normal unital positive right inverse.
    \end{enumerate}
    If these hold, then $E = \hat\beta^{-1} \circ \beta$ is a conditional expectation onto the multiplicative domain $\mult(\beta)$.
    Thus, $\beta$ factorizes into a conditional expectation part and an isomorphism part: 
    \begin{equation}\label{eq:surjective-factorization}
        \beta = \hat\beta\circ E.
    \end{equation}
    If $M,N$ are von Neumann algebras and $\beta$ is completely positive, then the equivalent properties guarantee that $\hat\beta$ is a *-isomorphism.
\end{lemma}
\begin{proof}
    For notational ease, set $D = \mult(\beta)$.
    The implication \ref{it:surjective3} $\Rightarrow$ \ref{it:surjective1} is clear.
    
    \ref{it:surjective1} $\Rightarrow$ \ref{it:surjective2}:
    Let $p\in N$ be a projection and let $h\in [0,1]_{M}$ be such that $\beta(h)=p$.
    By the Jordan-Schwarz inequality (see \cref{lem:multiplicative-dom}), we have $p = \beta(h) \ge \beta(h^2) \ge \beta(h)^2 = p^2 = p=\beta(h)$. 
    This implies $\beta(h)^2=\beta(h^2)$, and hence, $h\in D$.
    Since $p$ was arbitrary, we conclude that the J*-homomorphism $\hat\beta=\beta|_D$ is surjective.\footnote{Since $p$ was arbitrary, the range of the J*-homomorphism $\hat\beta = \beta|_D$ contains every projection of $N$. As $\hat\beta$ is normal, its range is an ultraweakly closed JBW*-subalgebra of $N$, and since a JBW*-algebra is generated by its projections, the surjectivity of $\hat\beta$ follows.}
    Since $\beta$ is faithful, $\hat\beta$ is faithful, and hence, injective.
    Therefore, $\hat\beta$ is a J*-isomorphism onto $N$.
    
    \ref{it:surjective2} $\Rightarrow$ \ref{it:surjective3}:
    Denote the embedding map $D\hookrightarrow M$ by $\iota$.
    We claim that the normal unital positive map $\gamma= \iota \circ \hat\beta^{-1} : N \to D \to M$ is a right inverse for $\beta$.
    Indeed, we have $(\beta\circ \gamma) \circ \hat\beta = \beta\circ \iota\circ\hat\beta^{-1}\circ \hat\beta = \beta\circ\iota = \hat\beta$. 
    Since $\hat\beta$ is surjective by assumption, this shows $\beta\circ\gamma=\id$.
    
    We show the last claim: Set $E = \hat\beta^{-1}\circ \beta : M \to D$. By definition, this is a normal unital positive map, which satisfies $E|_D = \hat\beta^{-1} \circ \beta|_D = \hat\beta^{-1}\circ\hat\beta = \id_D$.
    Thus, $E$ is a conditional expectation onto $D$, and we have $\hat\beta \circ E = \hat\beta\circ(\hat\beta^{-1}\circ \beta) = \beta$.
\end{proof}

\begin{lemma}\label{lem:order-interval}
    Let $\alpha : (N,\phi) \to (M,\omega)$ be a state-preserving normal unital positive map, where $\omega$ and $\phi$ are both faithful.
    Let $\beta: (M,\omega)\to (N,\phi)$ be the Petz dual of $\alpha$.
    The following are equivalent:
    \begin{enumerate}[(a)]
        \item 
        For every $0\le \phi_1\le\phi$, there is a $0\le \omega_1\le\omega$ such that $\phi_1=\omega_1\circ\alpha$;
        \item
        $\beta$ maps $[0,1]_M$ onto $[0,1]_N$.
    \end{enumerate}
\end{lemma}
\begin{proof}
    The first item can be rephrased by saying that the pre-adjoint map $\alpha_*$ maps $[0,\omega]$ onto $[0,\phi]$.
    Since $\Gamma_\omega$ is a order isomorphism between $[0,1]_M$ and $[0,\omega]$ and analogously for $\Gamma_\phi$, the order interval-surjectivity of $\alpha_{*}$ is equivalent to the order interval-surjectivity of $\beta = (\Gamma_\phi)^{-1}\circ\alpha_*\circ \Gamma_\omega$. 
\end{proof}

\begin{proof}[Proof of \cref{thm:lifting}]
    \ref{it:lifting1} $\Rightarrow$ \ref{it:lifting2}: Consider the ensemble $\{\phi_1, \phi - \phi_1\}$. 
    
    \ref{it:lifting2} $\Rightarrow$ \ref{it:lifting3} and \ref{it:lifting4}: 
    We first give a proof under the additional assumption that $\omega$ and $\phi$ are faithful.
    Let $\beta$ be the dual map of $\alpha : (N,\phi)\to (M,\omega)$. 
    Then $\alpha$ and $\beta$ are faithful.
    By \cref{lem:order-interval}, $\beta$ maps $[0,1]_M$ onto $[0,1]_N$.
    We denote its multiplicative domain by $D$.
    Applying \cref{lem:surjective} to $\beta$, it follows that $\hat\beta = \beta|_D$ is a J*-isomorphism onto $N$ and that $E = \hat\beta^{-1}\circ\beta$ is a normal conditional expectation of $M$ onto $D$ such that $\beta = \hat\beta \circ E$.
    We note that $\phi \circ \hat\beta = \omega |_D$ and therefore 
    \begin{align}
        \omega = \phi \circ \beta = \phi \circ \hat\beta \circ E = \omega|_D \circ E.
    \end{align}
    Therefore, we have a commutative diagram of state-preserving maps
    \begin{equation}
    \begin{tikzcd}
        (M,\omega) \arrow{rr}{\beta}\arrow{dr}{E} && (N,\phi) \\
        & (D,\omega|_D) \arrow{ur}{\hat\beta}
    \end{tikzcd}
    \end{equation}
    Since the states involved are all faithful, we can consider the dual maps.
    By \cref{lem:functor}, this preserves commutativity of diagrams.
   By definition, the dual map of $\beta$ is $\alpha$, and, by \cref{it:functor3} of \cref{lem:functor}, the dual map of $\hat\beta$ is the J*-isomorphism $\hat\beta^{-1}$ of $N$ onto $D$. The dual map of $E$ is the embedding $\iota: D \hookrightarrow M$.
    Thus, we arrive at the following commutative diagram of state-preserving maps
    \begin{equation}
    \begin{tikzcd}
        (M,\omega)  && (N,\phi) \arrow{ll}{\alpha}
        \arrow{dl}{\hat\beta^{-1}} \\
        & (D,\omega|_D) \arrow[hook]{ul}{\iota}
    \end{tikzcd}
    \end{equation}
    This shows that $\alpha = \iota\circ \hat\beta^{-1}$ is a J*-isomorphism onto the JBW*-subalgebra $D\subset M$, which admits a $\omega$-preserving conditional expectation $E$. 
    Moreover
    \begin{align}
    \beta \circ \alpha = \beta\circ \iota\circ\hat\beta^{-1} = \hat\beta \circ \hat\beta^{-1} =  \id_N.
    \end{align}
    
    The general case, where the states are not necessarily faithful, follows by applying the solution of the faithful case to the map $\hat\alpha = P_{\supp \omega} \circ \alpha|_{N_{\supp\phi}}$ between the support corners $N_{\supp\phi}$ and $M_{\supp\omega}$ of the two states.
    If $M, N$ are von Neumann algebras and if $\alpha$ is completely positive, the constructions above automatically ensure that $D$ is a von Neumann algebra, that $\beta$ is completely positive (see \cref{lem:functor}), and that $\hat\alpha$ is a *-isomorphism.

   \ref{it:lifting4} $\Rightarrow$ \ref{it:lifting3}: 
    Consider the unital positive map $\beta = \hat\alpha^{-1} \circ E \circ P_{\supp \omega} + (1-\supp\phi)\cdot \omega$.
    Then, since $P_{\supp\phi}(1-\supp\phi)=0$, $E\circ\hat\alpha = \hat\alpha$,  and $P_{\supp \phi}|_{N_{\supp\phi}} = \id_{N_{\supp \phi}}$, we have
    \begin{align}
      P_{\supp \phi}\circ  \beta \circ \alpha|_{N_{\supp\phi}} \nonumber
      &= P_{\supp \phi}\circ \hat\alpha^{-1} \circ E \circ P_{\supp\omega} \circ \alpha|_{N_{\supp\phi}} + 0 \nonumber\\ 
      &= \hat\alpha^{-1} \circ E \circ \hat\alpha \nonumber\\
      &= \hat\alpha^{-1} \circ \hat\alpha \nonumber\\
      &= \id_{N_{\supp\phi}}.
    \end{align}
    Moreover, since $\omega\circ \hat\alpha = \phi|_{N_{\supp \phi}}$, it follows that $\phi \circ \hat\alpha^{-1} = \omega|_L$. 
    Hence $\phi \circ \hat\alpha^{-1} \circ E = \omega \circ E = \omega|_{M_{\supp \omega}}$ and we get
    \begin{align}
        \phi \circ \beta 
        = \phi \circ \hat\alpha^{-1} \circ E \circ P_{\supp \omega} + \phi(1-\supp\phi)\,\omega
        = \omega|_{M_{\supp \omega}} \circ P_{\supp\omega} +0 = \omega.  
    \end{align}

    \ref{it:lifting3} $\Rightarrow$ \ref{it:lifting1}:  
    Set $\omega _x = \phi_x \circ \beta = \phi_x\circ P_{\supp\phi} \circ \beta$. 
    Then $\{\omega_x\}$ is an ensemble with average state $\omega$. Indeed,
    since $\beta$ is state-preserving, we have $\sum_x \omega_x = \phi \circ \beta = \omega$. 
    Since $\sum_x \omega_x \circ \alpha = \phi$, we have $\omega_x \circ \alpha \leq \phi$ and hence $\omega_x \circ \alpha = \omega_x \circ \alpha\circ P_{\supp \phi}$. 
    Therefore, using $P_{\supp \phi} \circ \beta\circ \alpha|_{N_{\supp \phi}} = \id_{N_{\supp \phi}}$, we find
    \begin{align}
        \omega_x \circ \alpha 
        = \omega_x \circ \alpha \circ P_{\supp \phi} 
        = \phi_x \circ (P_{\supp \phi}\circ \beta \circ \alpha) \circ P_{\supp \phi}  
        = \phi_x \circ P_{\supp\phi} = \phi_x. 
    \end{align}
\end{proof}

\section{Steering and the ensemble lifting property}
\label{sec:steering}

Having established the general lifting property for positive maps and its connection to conditional expectations, we now turn to the application to steering.
Recall that a POVM in a von Neumann algebra $M$ is a collection $\{a_x\}_{x\in X}$ of positive elements $0\le a_x$ such that $\sum_x a_x=1$.
We often suppress the index set $X$, which is always assumed to be finite or countably infinite set.

\begin{definition}\label{def:steering}
Let $A,B\subset B(\H)$ be commuting von Neumann algebras.
Let $\omega$ be a normal state on $B(\H)$.
We say that $B$ can steer an ensemble $\{\omega_{A,x}\}$ on $A$ relative to $\omega$ if there exists a POVM $\{b_x\} \subset B$ such that
\begin{align}
    \omega_{A,x} = \omega((\placeholder) b_x)|_A \qquad \text{for all $x\in X$}. 
\end{align}
If this holds for all ensembles on $A$ whose average state is $\omega_A$, we say that $B$ can steer $A$ relative to $\omega$.
\end{definition}

Note that whether steering is possible relative to $\omega$ depends only on the restriction of the state to the von Neumann algebra $A\vee B$ generated by $A$ and $B$.

As an application of the commutant Radon-Nikodym theorem, we find that steering is always possible if $A$ and $B$ are commutants of each other:

\begin{lemma}\label{lem:hd-implies-steering}
    Let $\Omega \in \H$ be a state vector, and let $A\subset B(\H)$ be a von Neumann algebra, then $A'$ can steer $A$ relative to $\Omega$.
\end{lemma}

\begin{proof}
    Denote by $p = [A\Omega] \in A'$ the projection onto the cyclic subspace $\K = p\H$ of $\Omega$ relative to $A$ and set $\pi(a) = pap = ap$. Then $(\pi,\K,\Omega)$ is the GNS triple of $\omega_A$.
    By the Radon-Nikodym theorem, there is a POVM $\{ \hat b_x \} \subset \pi(A)' = (Ap)' = p(A')p = pA'p \subset A'$ such that $\omega_{A,x} = \langle \Omega,(\cdot) \hat b_x \Omega\rangle$.
    Setting $b_x = \hat b_x + \omega_{A,x}(1)(1-p) \in A'$ we find that
    $\sum_x b_x = p + (1-p) = 1$ and,  since $(1-p)\Omega = 0$,
    \begin{equation*}
        \langle \Omega, a b_x \Omega \rangle = \langle\Omega, a \hat b_x \Omega\rangle  = \omega_{A,x}(a), \quad a\in A.\qedhere
    \end{equation*}
\end{proof}

As a direct consequence, we obtain the equivalence of steering and ensemble extension (this is \cref{introlem} in \cref{sec:intro}):

\begin{corollary}\label{cor:steer-iff-extend}
     Let $\Omega\in \H$ be a state vector and let $\{\omega_{A,x}\}$ be an ensemble on $A$ with $\sum_x\omega_{A,x}=\omega_A$.
    The following are equivalent:
    \begin{enumerate}[(a)]
	\item\label{it:steer-iff-extend1} $B$ can steer $\{\omega_{A,x}\}$ relative to $\Omega$.
	\item\label{it:steer-iff-extend2} 
    The ensemble $\{\omega_{A,x}\}$ can be extended to an ensemble on $B'$ whose average state is $\omega_{B'}$, i.e., there exists an ensemble $\{\omega_{B',x}\}$ with $\omega_{B',x}|_A = \omega_{A,x}$ and $\sum_x \omega_{B',x}= \omega_{B'}$.
    \end{enumerate}
\end{corollary}

\begin{proof}
    If $\{b_x\} \subset B$ is a POVM such that $\omega_{A,x}(a) = \ip\Omega{ab_x\Omega}$, $a\in A$, then an ensemble extension is given by $\omega_{B',x}(b') = \ip\Omega{b'b_x\Omega}$.
    Conversely, if $\{\omega_{B',x}\}$ is an ensemble extension, then, by \cref{lem:hd-implies-steering}, there is a POVM $\{b_x\} \subset B$, such that $\omega_{B',x}(b') = \ip\Omega{b'b_x\Omega}$, $b'\in B'$.
    Thus, $B$ can steer the ensemble $\{\omega_{B',x}\}$ on $B'$ relative to $\Omega$.
    Hence, $B$ can also steer the restricted ensemble $\{\omega_{A,x}\}$ on $A$ relative to $\Omega$.
\end{proof}

\subsection{Characterization of one-way steering}

In this section we characterize one-way steering. We begin by proving \cref{introthm:main}, which we restate here for convenience:

\begin{theorem}\label{thm:main}
    Let $A,B\subset B(\H)$ be commuting factors, let $\Omega\in\H$ be a state vector.
    The following are equivalent:
    \begin{enumerate}[(a)]
        \item\label{it:intromain1}$B$ can steer $A$ relative to $\Omega$;
        \item\label{it:intromain2} every ensemble on $A$ with average state $\omega_A$ extends to an ensemble on $B'$ with average state $\omega_{B'}$;
        \item\label{it:intromain3} there exists an $\Omega$-preserving normal conditional expectation $E:B'\to A$, i.e., $\omega_{B'}=\omega_A \circ E$.
    \end{enumerate}
	The equivalent conditions imply that every normal state $\psi_A$ on $A$ with $\supp\psi_A\lesssim \supp\omega_A$ has a purification $\Psi\in \H$ that allows $B$ to steer $A$ relative to $\Psi$.
\end{theorem}

We need the following lemma that allows us to extend corner-conditional expectations in inclusions of von Neumann algebras.
In the following, if $p \in B(\H)$ is a projection and $N\subset B(\H)$ is a von Neumann algebra, we set
\begin{align}
c_N(p) := \inf \{ e\in \proj(N) : e\geq p \}
\end{align}
and call $c_N(p)$ the \emph{cover} of $p$ in $N$. 
We call $z_N(p) := c_{Z(N)}(p) \in Z(N)$ the \emph{central cover} of $p$ in $N$. 
Note, that $z_{N}(p) = z_{N'}(p)$.
The central support of a normal state $\phi$ on a von Neumann algebra $M$ is central cover of the support $\supp\phi$ in M.

\begin{lemma}\label{lem:extending-CEs}
    Let $N\subset M$ be von Neumann algebras.  
    \begin{enumerate}[(i)]
        \item\label{it:extending-CEs1} Let $p \in N$ be a projection with central cover 1.
        Then, every normal conditional expectation $E:pMp\to pNp$ extends to a normal conditional expectation $F: M \to N$.
        \item\label{it:extending-CEs2} 
        Let $e \in N'\cap M$ be a projection with central cover $1$ in $N$, i.e., the only central projection $f\in N$ with $f\ge e$ is $f=1$.
        Then, every normal conditional expectation $E:eMe\to eN$ extends to a normal conditional expectation $F: M \to N$ in the sense that $eF(x) = E(x)$ for all $x \in eMe$.
    \end{enumerate}
\end{lemma}

\begin{proof}
    \ref{it:extending-CEs1}: Because the central cover of $p$ is $1$, we can construct, using Murray-von Neumann comparison theory, a mutually orthogonal family of projections $(p_n)_{n\in \NN}$ in $N$ with $\sum p_n =1$, $p_1=p$, and $p_n \lesssim p$ (cp.~\cite[Prop.~III.1.1.6]{blackadar_operator_2006}).
    For each $n\in\NN$, let $v_n$ be a partial isometry with $v_n^*v_n=p_n$ and $v_nv_n^* \le p$. 
    Note that $v_nxv_m^* \in pMp$ for all $x\in M$.
    Define
    \begin{equation*}
        F(x) = \sum_{n,m} v_n^* E(v_nxv_m^*) v_m.
    \end{equation*}
    Clearly $F$ is normal, $F|_{pMp} =E$, and $F^2=F$.
    Moreover, if $x\in N$, then $v_n x v_m^* \in pNp$ and we have $F(x) = \sum_{n,m} v_n^*v_nx v_m^*v_m = x$. Thus, $F$ is a normal conditional expectation onto $N$, which extends $E$.

    \ref{it:extending-CEs2}:
    Since the central cover of $e$ in $N$ is $1$, the cut-down (or induction) map $I_{e}: N \to eN$, $a\mapsto ea$, is a *-isomorphism \cite[Prop.~3.14]{stratila_lectures_2019}.
    We define $F$ as $F = I_{e}^{-1}\circ E \circ P_e$, where $P_e$ is the cut-down map $M \to eMe$.\footnote{Note that $P_{e}$ is an $N$-bimodule map.}
    By definition, $F$ is normal, unital and completely positive. 
    It satisfies $F|_N=\id$ since
    \begin{equation}
        F(a) 
        = I_{e}^{-1} (E(eae) )
        = I_{e}^{-1}(ea) = a, \quad a\in N.
    \end{equation}
     Finally, $eF(x) = I_{e}(F(x)) = E(exe)$ for every $x\in M$, i.e., $eF(x)=E(x)$ for $x\in eMe$, which establishes the claim.
\end{proof}

\begin{lemma}\label{lem:cut-down-iso}
    $N\subset M$ be von Neumann algebras let $q\in N$, $p\in M$ be projections with $p\le q$.
    Suppose that the restriction of the normal unital completely positive map $qNq \to pMp$, $a\mapsto pap$, is injective and multiplicative.
    Then $p = qe$ for some projection $e\in N'\cap M$. 
    Moreover, if $q$ has central cover $1$ in $N$, then the central cover of $p$ in $eN$ is $e$, and the central cover of $e$ in $N$ is $1$.
\end{lemma}
\begin{proof}
    Set $R=N'\cap M$ and note $q\in R'$.
    By assumption, $pep$ is a projection for every projection $e\in qNq$.
    Thus, $p$ commutes with every projection in $qNq$, and therefore with all of $qNq$. 
    Since $p\in qMq$ (because $p\leq q$), $p \in (qNq)'\cap qMq = qN'q \cap qMq = qR$.
    Since the cut-down $R \to qR$ is a normal *-homomorphism, the central cover $z_{R}(q)\in Z(R)$ of $q$ lets us restrict $R \ni r\mapsto qr q=qr$ to a *-isomorphism $z_{R}(q)R\cong qR$. 
    Thus, we obtain a unique element $f\in z_{R}(q)R$ such that $qf=p$, and we set $e = fz_{R}(q) + (1-z_{R}(q))\in R$. Clearly, we have $qe = qf = p$.
    
    Now suppose that $z_{N}(q)=1$.
    Let $c\in Z(N)$ be a projection with $c\ge p$. Then $(1-c)p=0$, so that $p (q(1-c))p = 0$. 
    Since the cut-down map $qMq\to pMp$, $a\mapsto pap$, is injective on $qNq$ and $q(1-c)\in qNq$, this implies $q(1-c)=0$, so that $c=1$ since $z_{N}(q)=1$ in $N$.
    Thus, the $N$-central cover of $p$ is $1$.
    Since $ep = p$, this implies that $p$ has $eN$-central cover $e$, and that $e$ has $N$-central cover $1$.
\end{proof}

We can now prove \cref{introprop:qi-takesaki}:

\begin{corollary}\label{prop:qi-takesaki}
    Let $N\subset M$ be von Neumann algebras. 
    Let $\omega$ be a normal state on $M$ and let $\phi=\omega|_N$ be its restriction to $N$. 
    Suppose $\phi$ has central support $1$.
    Then, the following are equivalent:
    \begin{enumerate}[(a)]
        \item\label{it:qi-takesaki1} for every $\phi_1\in N_*$ with $0\le\phi_1\le\phi$, there exists $\omega_1\in M_*$ with $0\le\omega_1\le \omega$ and $\omega_1|_N=\phi_1$;
        \item\label{it:qi-takesaki2} there is an $\omega$-preserving normal conditional expectation $E:M\to N$, i.e., $\omega=\phi\circ E$.
    \end{enumerate}
\end{corollary}

\begin{proof}
    As before, \ref{it:qi-takesaki2} $\Rightarrow$ \ref{it:qi-takesaki1} follows by setting $\omega_1= \phi_1\circ E$.
    
    \ref{it:qi-takesaki1} $\Rightarrow$ \ref{it:qi-takesaki2}:
    Set $s_M = \supp\omega$, $s_N=\supp\phi$, and note $s_M\le s_N$.
    We apply \cref{thm:lifting} to the inclusion map $\alpha= \iota:N\hookrightarrow M$.
    By assumption item \ref{it:lifting1} of \cref{thm:lifting} holds, and we may evaluate item \ref{it:lifting2} of the theorem.
    The map $\hat\alpha: s_NNs_N \to s_MMs_M$, $s_Nas_N\mapsto s_Mas_M$, is a multiplicative and injective normal completely positive map onto its range $s_MNs_M$ ($=L$ in the notation of \cref{thm:lifting}).
    By \cref{lem:cut-down-iso}, there is a projection $e\in N'\cap M$ such that
    \begin{equation}\label{eq:another-support-factorization}
        s_M=s_Ne.
    \end{equation}
    Since, by assumption, $s_N$ has central cover 1, the lemma also implies that $s_M$ has $eN$-central cover $e$, and $e$ has $N$-central cover $1$.
    Hence, we have $s_MNs_M = s_M(eN)s_M$, $s_M \in eN$, and, by \cref{thm:lifting}, there exists a normal conditional expectation
    \begin{equation}
        F: s_MMs_M = s_M(eM)s_M \to s_M(eN)s_M = s_MNs_M,
    \end{equation}
    preserving the cut-down of $\omega$ by $s_{M}$.
    We apply \cref{lem:extending-CEs} twice to extend $F$ to a conditional expectation of $M$ onto $N$:
    As noted, the central cover of $s_M$ in $eN$ is $e$.
    Thus, the first item of \cref{lem:extending-CEs}, extends $F$ to a normal conditional expectation $\hat F : eM \to eN$.
    Since $e$ has central cover $1$ in $N$, the second item of \cref{lem:extending-CEs}, gives a normal conditional expectation $E:M\to N$ such that $ e E(x) = \hat F(x)$, $x\in eMe$.
    By construction, $E$ preserves $\omega$.
\end{proof}

We can now prove \cref{thm:main}: 

\begin{proof}[Proof of \cref{introthm:main}]
    The equivalence \ref{it:intromain1} and \ref{it:intromain2} is shown in \cref{cor:steer-iff-extend}.
    
    \ref{it:intromain3} $\Rightarrow$ \ref{it:intromain2}: 
    From a given ensemble $\{\omega_{A,x}\}$ on $A$ with average state $\omega_A$, we get an ensemble $\{\omega_{B',x}\}$ on $B'$ with average state $\omega_{B'}$ via $\omega_{B',x}=\omega_{A,x}\circ E$.

    \ref{it:intromain2} $\Rightarrow$ \ref{it:intromain3}: 
    In particular, binary ensembles extend.
    Hence, for every $\phi_A\in A_*$ with $0\le \phi_A\le \omega_A$, there exists an extension $\phi_{B'}\in (B')_*$ with $0\le \phi_{B'}\le\omega_{B'}$.
    Thus, the claim follows from \cref{prop:qi-takesaki}.
    
    We show the final statement in the theorem. Let $s_X=\supp \omega_X$, $X=A,B'$, and note $s_A\ge s_{B'}$.
    Let $e=\supp E\in A'\cap B'$, and assume first that $\supp\psi_{A}\leq s_{A}$. 
    We note that $s_{B'} = s_{A}e = [B\Omega]$ is a cyclic projection in $B'$. Then, any projection $p\in B'$ with $p\leq s_{B'}$ is also cyclic in $B'$ (with generating vector $p\Omega$) \cite[Prop.~5.5.9]{KadisonRingrose1}. 
    By assumption, $(\supp\psi_{A})e$ is cyclic in $B'$, because $(\supp\psi_{A})e\leq s_{A}e$, and, therefore, $\psi_{A}\circ E$ is a vector state of $B'$ \cite[Prop.~7.2.7]{KadisonRingrose2}. Since cyclicity of projections is preserved under Murray-von Neumann equivalence, the claim follows.
\end{proof}

\begin{remark}\label{rem:binary}
    We note that the application of \cref{thm:lifting} in the proof of \cref{introthm:main} only needs the existence of ensemble extensions of arbitrary binary ensembles. 
    Hence, item \ref{it:intromain2} of \cref{introthm:main} may be weakened to only the case of binary ensembles. 
\end{remark}

We now make use of \cref{thm:main} to study steering for mixed states in the irreducible case, i.e., when $A\vee B = B(\H)$. 
The result shows that, in the irreducible setting, steering in mixed states always reduces to steering in pure states.

\begin{proposition}\label{prop:mixed-one-way}
    Let $A,B$ be commuting factors on $\H$ with $A\vee B =B(\H)$, let $\omega$ be a normal state on $B(\H)$.
    Let $p = \supp(\omega_B) \in B$.
    The following are equivalent:
    \begin{enumerate}[(a)]
        \item\label{it:mixed-one-way1} $B$ can steer $A$ relative to $\omega$;
        \item\label{it:mixed-one-way2} 
        There is an $\omega$-preserving normal conditional expectation $E:B'\to A$ and there is a tensor product decomposition $p\H=\H_0\ox\H_1$ such that 
        $B'p\subset  B(\H_0) \ox 1$ and $\omega = \omega_0\ox\omega_1$ on $B(p\H)$ with $\omega_0$ pure. 
    \end{enumerate}
\end{proposition}

The following lemma is inspired by an argument in \cite{shiFusionRulesEntanglement2020}.

\begin{lemma}\label{lem:irreducible-up-to-type-I-CE}
    Let $N\subset M$ be an irreducible subfactor.
    Let $F: M\otimes B(\K) \to N\otimes 1$ be a normal conditional expectation. Then, there is a normal conditional expectation $E:M\to N$ and a normal state $\sigma$ on $B(\K)$ such that
    \begin{equation}
        F = E \otimes \sigma.
    \end{equation}
\end{lemma}
\begin{proof}
    Let $\sigma$ be the state defined by $\sigma(y) = F(1\ox y)$, $y\in B(\K)$, and let $E:M\to N$ be the conditional expectation defined by $E(x) =F(x\otimes 1)$, $x\in M$.
    Given $0\le y\in B(\K)$, consider the normal completely positive map $F_y=F(\placeholder\otimes y) : M \to N$.
    The bimodule property of $F$ implies that $F_y$ is a bimodule map, i.e., $F_y(axb)=aF_y(x)b$, $x\in M$, $a,b\in N$, hence an operator-valued weight.
    Irreducibility ensures that all operator-valued weights $M\to N$ are proportional \cite[Cor.~12.13]{stratila_modular_2020}.
    Thus, $F_y\propto E$.
    Evaluating on $1\in M$, gives $F_y = \sigma(y)E$.
    Hence, $F(x\otimes y) = F_y(x) = \sigma(y)E(x)$ for all $x\in M$, $0\le y\in B(\K)$.
    By linearity, the positivity assumption on $y$ can be dropped.
\end{proof}

\begin{lemma}\label{lem:uncorrelated-purification}
    Let $M$ be a factor on a Hilbert space $\H$ and let $\omega$ be a normal state on $B(\H)$ whose marginal on $M$ is faithful.
    The following are equivalent:
    \begin{enumerate}[(a)]
        \item \label{it:uncorrelated-purification1}
        There exists a purification $\Omega \in \H\ox \K$ of $\omega$ that is uncorrelated between $M'$ and the purifying system $\K$, i.e., for $a\in M$, $b\in M'$, $c\in B(\K)$, one has
        \begin{equation}
            \ip\Omega{(a\ox 1)\Omega} = \omega(a), \qquad \ip\Omega{(b\ox c)\Omega} = \omega(b)\,\ip\Omega{(1\ox c)\Omega}.
        \end{equation}
        \item\label{it:uncorrelated-purification2}
        There is a tensor product decomposition $\H =\H_0\ox \H_1$ such that $M'\subset B(\H_0)\ox 1$ and $\omega = \omega_0\ox\omega_1$ with $\omega_0$ a pure state on $B(\H_0)$.
    \end{enumerate}
\end{lemma}

\begin{proof}
    \ref{it:uncorrelated-purification1} $\Rightarrow$ \ref{it:uncorrelated-purification2}:
    Let $\sigma$ be the normal state induced by $\Omega$ on $\K$. 
    Since $\omega_M$ is faithful on $M$, $\Omega$ is cyclic for $M'\otimes B(\K)$.
    Therefore, the given representation is a GNS representation of the product state $\omega_{M'}\ox\sigma$ on $M'\ox B(\K)$.
    Consider GNS representation the $(\H_{\sigma}, \pi_{\sigma},\Omega_{\sigma}) $ of $(B(\K),\sigma)$. Using the fact that $B(\K)$ is of type I, we take
    $(\H_{\sigma}, \pi_{\sigma}, \Omega_{\sigma})$ to be of the form 
    \begin{align}
        \H_{\sigma} = \H_{1} \otimes \K, \qquad \H_{1}=\supp(\sigma)\K, \qquad \pi_{\sigma}(c) = 1\otimes c, \quad c\in B(\K).
    \end{align}
    Given the GNS representation $(\H_0,\pi_0,\Omega_0)$ of $(M',\omega_{M'})$ and the uniqueness of the GNS representation of $\omega_{M'}\otimes\sigma$, there is a canonical unitary $U:\H\ox\K \to \H_0 \ox \H_1\ox \K$ defined by
    \begin{equation}
        U \Omega = \Omega_0 \ox \Omega_1, \qquad U (b\ox c) U^*=  \pi_0(b)\ox 1_{\H_1}\ox c , \qquad b\in M',\ c\in B(\K).
    \end{equation}
    In particular, $U (1_\H \ox c)U^* = 1_{\H_0\ox\H_1} \ox c$, $c\in B(\K)$, so that $U$ must be of the form $U = V \ox 1_\K$ for a unitary $V:\H\to\H_0\ox\H_1$.
    Suppressing the unitary $V$, we have $\H=\H_0\ox\H_1$ and $M' = \pi_0(M') \ox 1_{\H_1} \subset B(\H_0)\ox1_{\H_1}$, as required.
    Since $\Omega = \Omega_0\ox \Omega_1$, we have 
    \begin{align}
        \omega 
        =  \langle\Omega_0,\placeholder \Omega_0\rangle \ox \omega_1,
    \end{align}
    with $\omega_1 = \langle\Omega_1,\placeholder \Omega_1\rangle|_{B(\H_1)\ox 1}$. Thus $\omega$ factorizes and its reduction to $B(\H_0)\ox1$ is pure.

    \ref{it:uncorrelated-purification2} $\Rightarrow$ \ref{it:uncorrelated-purification1}:
    Similar to preceding step, we purify the state $\omega_{1}$ on $B(\H)$ by a vector $\Omega_1\in \H_1\ox \K$ for some Hilbert space $\K$, and we let $\Omega_0\in \H_0$ be a state vector implementing the pure state $\omega_0$.
    Let $\Omega=\Omega_0\ox\Omega_1 \in \H_0\ox(\H_1\ox\K) = \H\ox\K$.
    Then $\Omega$ purifies $\omega$, and is uncorrelated between $M'\ox 1_{\K} \subset  B(\H_0)\ox 1_{\H_1\ox \K}$ and $1_\H\ox B(\K) \subset 1_{\H_0}\ox B(\H_1\ox \K)$.
\end{proof}

\begin{lemma}\label{lem:steering-support-B}
     Let $A,B$ be commuting factors on $\H$, $\omega$ be a normal state on $\B(\H)$, and set $p = \supp(\omega_B)\in B$.  
    Then the following are equivalent:
    \begin{enumerate}[(a)]
        \item \label{it:steering-support-B1} $B$ can steer $A$ relative to $\omega$ on $\H$.
        \item \label{it:steering-support-B2} $pBp$ can steer $Ap$ relative to $\omega$ on $p\H$. 
    \end{enumerate}
\end{lemma}

\begin{proof}
The claim is immediate from the following two observations: 1) For $p \geq \supp(\omega)$, we have $\omega = p\omega p$ and $A\cong Ap$ because $p\in A'$. This gives a bijective correspondence between normal positive linear functionals $\psi_{A}$ on $A$ and those on $Ap$ by $\psi_{A}(a):= \psi_{Ap}(ap)$, and we have $\omega_{Ap} = (p\omega p)_{Ap} = (p\omega p)_{A} = \omega_{A}$. 2) Any POVM $\{pb_x p\}$ in $pBp$ can be completed to a POVM $\{b_x\}$ in $B$ by adding a POVM $\{b^{c}_{x}\}$ in $(1-p)B(1-p)$, i.e., $b_{x} = pb_{x}p+b^{c}_{x}$.
\end{proof}

The cut-down by $p=\supp(\omega_B)$ also leaves the steering capability of $A$ unchanged:

\begin{lemma}\label{lem:steering-support-B-dual}
    Let $A,B$ be commuting factors on $\H$, let $\omega$ be a normal state on $\B(\H)$, and set $p=\supp(\omega_B)\in B$.
    Then $A$ can steer $B$ relative to $\omega$ on $\H$ if and only if $Ap$ can steer $pBp$ relative to $\omega$ on $p\H$.
\end{lemma}
\begin{proof}
    Since $\supp(\omega_B)=p$, restriction gives a bijection between ensembles on $B$ with average state $\omega_B$ and ensembles on $pBp$ with average state $\omega_{pBp}$.
    Since $p\in B\subseteq A'$ and $A$ is a factor, the cut-down $A\to Ap$ is a normal $*$-isomorphism, so $\{a_x\}\mapsto\{a_xp\}$ is a bijection between POVMs in $A$ and POVMs in $Ap$.
    Finally, $p\ge \supp\omega$ gives $\omega = p\omega p$ and hence, for $a\in A$ and $b\in B$,
    \begin{align}
        \omega(b\,a_x) = \omega\big((pbp)(a_xp)\big),
    \end{align}
    so the two steering conditions correspond to each other under these bijections.
\end{proof}

\begin{proof}[Proof of \cref{prop:mixed-one-way}]
    \ref{it:mixed-one-way1} $\Rightarrow$ \ref{it:mixed-one-way2}:
    Let $\K$ be a Hilbert space and let $\Omega\in \H\ox\K$ be a state vector purifying the state $\omega$ on $\B(\H)$.
    For $M\subset B(\H\ox\K)$, denote by $\omega_M$ the restriction of $\ip\Omega{(\placeholder)\Omega}$. We also write $\omega_{B(\H)}$ instead of $\omega$ for the state on $B(\H)$ to avoid confusion with the pure state induced by $\Omega$ on $B(\H\ox\K)$.
    Set $\hat A=A\ox1$, $\hat B= B\ox 1$.
    By Definition \ref{def:steering}, $\hat B$ can steer $\hat A$ relative to $\Omega$ if and only if $B$ can steer $A$ relative to $\omega_{B(\H)}$.

    Since $\hat B$ can steer $\hat A$ relative to $\Omega$, there exists an $\Omega$-preserving conditional expectation $F:\hat B' = B'\ox B(\K) \to \hat A=A\ox 1$.
    By \cref{lem:irreducible-up-to-type-I-CE}, there is a normal state $\sigma$ on $B(\K)$ such that $F = E\ox \sigma$ with $E:B'\to A$ an $\omega_{B(\H)}$-preserving conditional expectation, showing the first part of \ref{it:mixed-one-way2}.
    $F$ being $\Omega$-preserving implies $\sigma = \omega_{B(\K)}$ and
    \begin{equation}
        \omega_{B' \ox B(\K)} = \omega_{B'} \ox \omega_{B(\K)}.
    \end{equation}
    Applying \cref{lem:uncorrelated-purification} to $M= pBp$ on $p\H$implies the claim.

    \ref{it:mixed-one-way2} $\Rightarrow$ \ref{it:mixed-one-way1}:
    By \cref{lem:steering-support-B}, $B$ can steer $A$ relative to $\omega$ if and only if $pBp$ can steer $Ap$ relative to the cut-down state $\omega_{B(p\H)}=p\omega p$. 
    Since $p$ commutes with the factors $B'$ and $A$, we have isomorphisms $B'p\to B'$ and $A\to A p$. The given $\omega_{B(\H)}$-preserving conditional expectation $E$ induces an $\omega_{B(p\H)}$-preserving conditional expectation $E_p: B'p \to Ap$ acting as $E_p(bp)=E(b)p$ for $b\in B'$.
    Now, let $\Omega_1\in \H_1\ox \K$ be a purification of $\omega_1$, let $\Omega_0 \in \H_0$ be the vector implementing $\omega_0$ and write $\sigma = \ip{\Omega_1}{\placeholder \Omega_1}|_{B(\K)}$. Set $F_p = E_p \otimes \sigma$ and $\hat B_p = pBp \ox 1,\hat A_p = Ap\ox 1 \subset B(p\H\ox\K)$. 
    Then $(\hat B_p)' = B'p \ox B(\K)$ and 
    \begin{align}
        \ip{\Omega}{\placeholder \Omega}|_{\hat B_p'} = \omega_{B'p} \ox \sigma.
    \end{align}
    Thus, $F_p$ is an $\Omega_0 \ox \Omega_1$-preserving conditional expectation $\hat B'_p \to \hat A_p$, which implies \cref{it:mixed-one-way1} by \cref{lem:steering-support-B} and Theorem \ref{thm:main}.
\end{proof}

\subsection{Two-way steering}\label{sec:two-way}

In the following we assume that $A\subset B'$ is an irreducible subfactor inclusion. The question we ask is: If $B$ can steer $A$ relative to a state $\omega$ on $B(\H)$, can $A$ also steer $B$ relative to $\omega$?
The following proposition shows that this is in general not the case. In fact, we see that two-way steering relative to the same state $\omega$ is true if and only if Haag duality holds and the state $\omega$ is pure.

\begin{proposition}\label{prop:asymmetry}
    Let $A,B\subset B(\H)$ be commuting factors with $A\vee B = B(\H)$ and let $\omega$ be a normal state on $B(\H)$.
    Then the following are equivalent:
    \begin{enumerate}[(a)]
        \item\label{it:asymmetry1} $A$ and $B$ can steer each other relative to $\omega$;
        \item\label{it:asymmetry2} $A=B'$ and $\omega$ is a pure state.
    \end{enumerate}
\end{proposition}

The proof of the proposition requires the following elementary observation:
Suppose $E:M\to N$ is a normal conditional expectation and $\omega_N$ a normal state on $N$. 
Then 
\begin{align}\label{eq:support-factorization}
    \supp(\omega_N \circ E) = \supp(\omega_N)\cdot\supp E.
\end{align}
To see this, write $s=\supp(\omega_{N}\circ E), s_{N} = \supp(\omega_N)$, and $e=\supp E$. 
Since $\omega_N(E(s)) = 1$, we have $E(s) \geq s_N$. 
Similarly, we have $e\geq s$ since $\omega_{N}(E(e))=1$.
The bimodule property of $E$ implies $e\in N'\cap M$ and $\omega_N(E(es_N)) = \omega_{N}(E(e)s_{N}) = 1$, hence $s \leq e s_N \leq s_N$. Since $E$ is a positive map, this implies $E(s) \leq s_N$ and hence $E(s) = s_N$.
Since $ese=s$, we thus find $E(es_N -s) = E(e(es_n - s)e) = 0$.
Since $E$ is faithful on $eMe$ we find $s = e s_N$.

\begin{proof}
    The implication \ref{it:asymmetry2} $\Rightarrow$ \ref{it:asymmetry1} is shown in \cref{lem:hd-implies-steering}.

    \ref{it:asymmetry1} $\Rightarrow$ \ref{it:asymmetry2}: 
    We first prove the case where $\omega= \ip{\Omega}{\placeholder \Omega}$ is a pure state.
    By the stated assumptions, there exists an $\omega_{B'}$-preserving conditional expectation $E:B'\to A$ and an $\omega_{A'}$-preserving conditional expectation $F:A'\to B$. 
    We write $s_X = \supp \omega_X = [X'\Omega]$. 
    By \eqref{eq:cyclicity}, we have
    \begin{equation}\label{eq:corners1}
        s_{A'} A = s_{A'} E(B') = s_{A'} B' s_{A'}.
    \end{equation}
    Next, note that the conditional expectation $F$ is faithful. Indeed, by the bimodule property, the support projection of a conditional expectation is an element of the relative commutant, which is trivial $A'\cap B' =\CC$.
    Thus, we have $s_{A'} = \supp(\omega_{B}\circ F) = \supp(\omega_B) \cdot \supp F = s_B$, giving $s_{A'}=s_B\in B.$
    Therefore, \eqref{eq:corners1} gives
    \begin{equation}\label{eq:corners2}
        s_{B} A = s_{B} B',
    \end{equation}
    which implies $A = B'$ since $A$ and $B'$ are factors.

    Now let $\omega$ be a general normal state on $B(\H)$.
    We write $p = \supp(\omega_{B})$.
    By \cref{prop:mixed-one-way}, we have $p \H = \H_0 \ox \H_1$ with 
    \begin{align}
       A p = A_0 \ox 1,\qquad pBp = B_0 \ox B(\H_1),\qquad A_0\vee B_0 = B(\H_0),\qquad p\omega p = \omega_0 \ox \omega_1,
    \end{align}
    with $\omega_0 = \ip{\Omega_0}{\placeholder \Omega_0}$ for some $\Omega_0 \in \H_0$. 
    Since $p\omega p$ is faithful on $p B p$, it follows that $\omega_1$ is faithful on $B(\H_1)$.
    Thus, either $\dim \H_1 = 1$ or there is an ensemble $\{\omega_{1,x}\}$ with average state $\omega_1$ that is non-trivial, where we say that an ensemble in trivial if each of its elements is proportional to the average state. 
    By assumption, $A$ can steer $B$, which implies that $A p = A_0 \ox 1$ can steer $p B p = B_0 \ox B(\H_1)$ by \cref{lem:steering-support-B-dual}. 
    Since $p \omega p$ is a product state, every ensemble that can be steered by $A p \subset B(\H_0)\ox1$ must restrict to a trivial ensemble on $B(\H_1)$. 
    Hence, $\dim \H_1 = 1$, so that $p \H = \H_0$, $Ap = A_0$, $p B p = B_0$, $A_0\vee B_0 = B(\H_0)$, and $p \Omega_0 = \Omega_0$.
    It follows that $A_0$ and $B_0$ can steer each other relative to $\Omega_0$. 
    Hence, the solution to the pure state case above implies $Ap = A_0 = B_0' = (p B p)' = B' p$.
    Since $p\in B\subseteq A'$, we conclude $A=B'$. 
\end{proof}

A more nuanced picture emerges once we allow $A$ and $B$ to steer each other relative to possibly distinct states, in which case a clear connection to the (Jones-Kosaki-Longo) index $[B':A]$ emerges \cite{jonesIndexSubfactors1983a,kosakiExtensionJonesTheory1986,longoIndexSubfactorsStatistics1989a,longoIndexSubfactorsStatistics1990,kosakiRemarkMinimalIndex1992}.
The crucial fact that we need is that for an irreducible subfactor inclusion $N\subset M$ the index $[M:N]$ is finite if and only if there exist (necessarily unique) normal faithful conditional expectations $E:M\to N$ and $E':N' \to M'$ (this is an immediate consequence of Haagerup's Cocyle Radon-Nikodym theorem for operator-valued weights \cite[Thm.~6.5]{haagerup_operator_1979-2}).\footnote{For an irreducible inclusion, if $[M:N]=\infty$ and a normal conditional expectation $E:M\to N$ exists, then no normal conditional expectation exists from $N'$ to $M'$ (this is also a direct consequence of \cite[Thm.~6.5]{haagerup_operator_1979-2})}.

\begin{proposition}\label{prop:index}
    Let $A,B\subset B(\H)$ be commuting factors with $A\vee B=B(\H)$. Then the following are equivalent:
    \begin{enumerate}[(a)]
        \item\label{it:index1} $[B':A] < \infty$;
        \item\label{it:index2} $B$ can steer $A$ relative to a state vector $\Psi \in \H$ and $A$ can steer $B$ relative to a state vector $\Phi \in \H$;
        \item\label{it:index3} $B$ can steer $A$ relative to normal state $\psi$ on $B(\H)$ and $A$ can steer $B$ relative to a normal state $\phi$ on $B(\H)$;
    \end{enumerate}
    If the equivalent conditions hold, we can have $\Psi = \Phi$ if and only if $A=B'$.
\end{proposition}

To prove the proposition we use the following lemma:
\begin{lemma}\label{lem:purification}
    Let $N\subset M$ be an irreducible subfactor inclusion with normal faithful conditional expectation $E:M\to N$ represented on a Hilbert space $\H$. 
    Then there exists a state vector $\Phi \in \H$ whose marginal states fulfill $\varphi_M = \varphi_N\circ E$.
\end{lemma}

\begin{proof}
    If $M$ has type I, then $M=N$ and $E=\id$, so that every state vector $\Phi \in \H$ works.
    If $M$ has type II with normal semifinite faithful trace $\tau$, then a normal state $\varphi$ on $M$ has a representation as a vector state if and only if $\dim_M \H_\varphi \leq \dim_M\H$, where $\H_\varphi$ is the GNS representation of $\varphi$.
    We can phrase this condition as $\tau(\supp \varphi)\leq \dim_M \H$.\footnote{This follows from viewing $H_\varphi$ as $L^2(M)\supp(\varphi)$ and using $\dim_M(L^2(M)e) = \tau(e)$ for any projection $e\in M$ (and similarly for $\H$).}
    Now consider a finite projection $e\in N$ with $\tau(e)\leq \dim_M\H$ and the normal state $\varphi_N = \tau(e\placeholder)/\tau(e)$ on $N$.
    Set $\varphi = \varphi_N \circ E$.
    Then, $\varphi$ has a vector representative $\Phi\in \H$ because, by \eqref{eq:support-factorization}, $\supp\varphi = \supp(\varphi_N) \cdot (\supp E) = e$.
    If $M$ has type III, then $M$ is in standard form and every normal state $\varphi$ on $M$ has a vector state representative $\Phi\in\H$. We can thus pick any normal state on $\varphi_N$ on $N$ and consider the invariant state $\varphi = \varphi_N \circ E$. 
\end{proof}

\begin{proof}[Proof of \cref{prop:index}]
    \ref{it:index2} $\Rightarrow$ \ref{it:index3} is clear.
    \ref{it:index3} $\Rightarrow$ \ref{it:index1}: 
    By \cref{prop:mixed-one-way}, there exist conditional expectations $E:B'\to A$ and $E':A'\to B$.
    By the irreducibility of the inclusion $A\subset B'$, all operator-valued weights of $B'$ onto $A$ are proportional \cite{haagerup_operator_1979-1,haagerup_operator_1979-2}.
    Hence, one has $E^{-1}\propto E'$, where $E^{-1}$ denotes the operator-valued weight dual to $E$ \cite{kosakiExtensionJonesTheory1986}.
    Hence, one has $E^{-1}(1)<\oo$, i.e., the index is finite \cite{kosakiExtensionJonesTheory1986}.

    \ref{it:index1} $\Rightarrow$ \ref{it:index2}:
    If the index is finite, there exist conditional expectations $E: B'\to A$ and $E': A'\to B$. 
    By \cref{lem:purification} there exist state vectors $\Phi,\Psi\in \H$ whose induced states are invariant under the given conditional expectations. 

    The statement that $\Psi = \Phi$ if and only if $A=B'$ is the content of \cref{prop:asymmetry}.
\end{proof}

\appendix

\section{Appendix: Kadison-Schwarz inequality for JB-algebras}

In this appendix we provide a short, self-contained proof of the Kadison-Schwarz inequality for JB-algebras, because we could not find an explicit proof in the literature.
JB-algebras are norm-complete, real Jordan algebras, and can be thought of as the self-adjoint, Jordan-algebraic version of abstract C* algebras. In particular, JBW-algebras are JB-algebras.
We refer to \cite{hanche-olsen_jordan_1984} for details.
The generalization to the complex case is contained in \cref{lem:multiplicative-dom}.
In the following, let $N$ and $M$ be JB-algebras.

\begin{proposition}[Kadison-Schwarz inequality for JB-algebras]\label{prop:kadison-schwarz}
    If $\varphi: N \to M$ is a unital positive map, then
    \begin{align}
        \varphi(a^2) \geq \varphi(a)^2,\quad a\in N. 
    \end{align}
\end{proposition}

The proof has four main ingredients. 
The first is that positivity is an intrinsic property:
\begin{lemma}\label{lem:intrinsic-positivity}
    Let $N_0\subset N$ be a JB-subalgebra containing $1$ and $x\in N_0$. Then $x\geq 0$ in $N_0$ if and only if $x\geq 0$ in $N$.
\end{lemma}
\begin{proof}
    The spectrum $\sigma(x)$ is defined relative to the associative subalgebra $C(x)\subset N_0$, generated by $x$. Since $x\geq 0$ is equivalent to $\sigma(x)\subset[0,\infty)$, positivity is independent of the ambient algebra. 
\end{proof}

The second is the \emph{Shirshov-Cohn} theorem, stating that a JB-algebra generated by two elements (and the unit) is automatically a JC-algebra, i.e., isomorphic to a norm-closed Jordan algebra in $B(\H)_{sa}$ with product $a\circ b = \frac{1}{2}(ab+ba)$ (see, for example, \cite[Thm. 7.2.5]{hanche-olsen_jordan_1984}). 
The third ingredient is the following observation. For  $x\in N$ set
\begin{align}
    U_x(b) =  2 x \circ (x\circ b) - x^2 \circ b.
\end{align}
In the case of a JC-algebra (and hence in the case of C*-algebras) this simply amounts to $U_x(b) = xbx$. 
\begin{lemma}\label{lem:app:U-positive}
    Let $0\leq b \in N$. Then $U_x(b)\geq 0$ for all $x\in N$. 
\end{lemma}
\begin{proof}
    Let $N_0$ be the JB-subalgebra generated by $x,b,1$. By the Shirshov-Cohn theorem we may assume $N_0\subset B(\H)_{sa}$. 
    By \cref{lem:intrinsic-positivity} it suffices to show $U_{x}(b) \geq 0$ when viewed as element of $B(\H)_{sa}$. 
    But then $U_{x}(b) = xbx$.
    Since $x=x^*$ and $b\geq 0$ it follows that $U_{x}(b) \geq 0$. 
\end{proof}

The final ingredient is an algebraic identity which only uses bilinearity of the Jordan product and the fact that $x\circ 1 = x$. 

\begin{lemma}\label{lem:algebra}
    Let $b_1,\ldots,b_n \in N$ satisfy $\sum_i b_i  =1$, let $\lambda_1,\ldots,\lambda_n \in \RR$ and set
    \begin{align}
        b = \sum_{i=1}^n \lambda_i b_i, \quad c = \sum_{i=1}^n \lambda_i^2 b_i. 
    \end{align}
    Then $c - b^2 = \sum_{i=1}^n U_{\lambda_i 1 - b}(b_i)$. 
\end{lemma}
\begin{proof}
    Set $x_i = \lambda_i 1 - b$. Direct computation shows that
    \begin{align}
        U_{x_i}(b_i) = 2 x_i \circ (x_i \circ b_i) - x_i^2 \circ b_i = \lambda_i^2 b_i - 2 \lambda_i (b\circ b_i) + 2 b \circ (b\circ b_i) - b^2 \circ b_i.
    \end{align}
    Summing over $i$ yields $\sum_i U_{x_i}(b_i) = c - b^2$. 
\end{proof}

\begin{corollary}\label{cor:kadison}
    Let $\varphi: N \to M$ be a unital positive map. Let $e_1,\ldots,e_n \in N$ be orthogonal projections with $\sum_i e_i = 1$, $\lambda_1,\ldots,\lambda_n \in \RR$, and put $a = \sum_i \lambda_i e_i$. Then $\varphi(a)^2 \leq \varphi(a^2)$.
\end{corollary}
\begin{proof}
    Set $b_i = \varphi(e_i) \geq 0$. Since $a^2 = \sum_i \lambda_i^2 e_i$ we have $b:= \varphi(a) = \sum_i \lambda_i b_i$ and $c:= \varphi(a^2) = \sum_i \lambda_i^2 b_i$. By \cref{lem:algebra} we have
    \begin{align}
        \varphi(a^2) - \varphi(a)^2 = c - b^2 = \sum_{i=1}^n U_{\lambda_i 1 - b}(b_i).
    \end{align}
    Since $b_i \geq 0$, we find that $\varphi(a^2) - \varphi(a)^2 \geq 0$ by \cref{lem:app:U-positive}.
\end{proof}

\begin{proof}[Proof of \cref{prop:kadison-schwarz}]
    We first prove the statement for JBW-algebras and then lift it to the case of JB-algebras. So let $N,M$ be JBW-algebras. 
    Fix $a\in N$ and consider the JBW-subalgebra $W(a)$ generated by $a$ and $1$. For any $\eps>0$ there are real numbers $\lambda_1,\ldots,\lambda_n$ and orthogonal projections $e_1,\ldots,e_n \in W(a)$ such that $\norm{a - a_\eps}<\eps$, where $a_\eps = \sum_i \lambda_i e_i$ (see, e.g., \cite[Prop.~4.2.3]{hanche-olsen_jordan_1984}). 
    Then $\norm{a^2 - a_\eps^2} \leq \norm{a-a_\eps}\norm{a+a_\eps} \leq \eps (2\norm{a}+\eps)$. 
    Choosing a null-sequence $\epsilon_n$ and setting $a_n := a_{\epsilon_n}$ we find (since $\varphi$ is norm-bounded)
    \begin{align}
        \varphi(a_n^2) - \varphi(a_n)^2 \longrightarrow \varphi(a^2)-\varphi(a)^2, \qquad n\to\infty,
    \end{align}
    in norm. Since $\varphi(a_n^2) - \varphi(a_n)^2 \geq 0$ by \cref{cor:kadison} and since the cone of positive operators is norm-closed, the result for JBW-algebras follows.    

    Now let $N,M$ be JB-algebras. Then the biduals $N^{**}, M^{**}$ are JBW-algebras containing $N$ and $M$ as subalgebras, respectively \cite[Thm.~4.4.3]{hanche-olsen_jordan_1984}. 
    The bidual map $\varphi^{**}: N^{**} \to M^{**}$ is unital and positive. Since positivity is intrinsic, we can apply our result for JBW-algebras and conclude that it holds for JB-algebras. 
\end{proof}

\begin{remark}
    Using the Kadison-Schwarz inequality for JB-algebras, one sees that the Jordan-Schwarz inequality \eqref{eq:JS-inequality} holds for positive unital maps between JB*-algebras (the same proof as in \cref{lem:multiplicative-dom} applies).
    As a consequence, \cref{lem:multiplicative-dom} generalizes fully to JB or JB*-algebra.
\end{remark}

\clearpage

\printbibliography

\end{document}